\documentclass[a4paper,11pt]{article}
\usepackage{jheppub}

\usepackage{graphicx}
\usepackage{tabularx}
\usepackage[usenames,dvipsnames]{color}

\usepackage{physics}

\usepackage{amsmath}
\usepackage{bbm}
\usepackage{amsfonts}
\usepackage{amssymb}
\usepackage{latexsym}
\usepackage{graphicx}
\usepackage[english]{babel}
\usepackage{multirow}
\usepackage{float}
\usepackage{url}
\usepackage{slashed}
\usepackage{xcolor} 
\usepackage{array}
\usepackage{cases}
\usepackage{paralist}

\usepackage{amssymb,amsfonts,amsmath,cancel,multirow,bm}

\title{Direct Detection and Cosmological Constraints of Dark Matter with Dark Dipoles}

\author[a,b]{Takumi Kuwahara,}
\author[c,d]{Jun-Chen Wang,}
\author[d]{Shu-Run Yuan}

\affiliation[a]{Center for Theoretical Physics and College of Physics, Jilin University, Changchun, 130012, China}
\affiliation[b]{Center for High Energy Physics, Peking University, Beijing 100871, China}
\affiliation[c]{Department of Physics and Jockey Club Institute for Advanced Study,\\
The Hong Kong University of Science and Technology,
Hong Kong S.A.R., China}
\affiliation[d]{School of Physics, Peking University, Beijing 100871, China}

\abstract{
    We study a fermionic dark matter candidate that couples to the standard model particles exclusively through electric and magnetic dipole operators mediated by a massive dark photon. 
    Such dipole portals naturally arise in dark sectors where the dark matter is neutral under a hidden $U(1)_D$, and they lead to phenomenology distinct from conventional vector-current interactions. 
    We consider the direct-detection signals arising from dark matter-nucleus scattering including the Migdal effect, dark matter-electron scattering, and semiconductor targets, which allow sensitivity to sub-GeV dark matter masses, together with the cosmological bounds from such as thermal relic abundance, cosmic microwave background, big-bang nucleosynthesis, and cosmic-rays.
    We find that the dark dipole coupling can be largely constrained by direct detection (in particular, electric dipole coupling). 
    However, the cosmological observations have already constrained most of the parameter space, in particular for magnetic dipole interactions of $U(1)_D$ for sub-GeV dark matter. 
    For the dark matter mass below 10\,MeV, the semiconductor (in particular, using skipper-CCD) experiments can play a crucial role in probing the dark dipole interactions: future low-threshold experiments utilizing the semiconductor targets can further extend the constraints.
    Our results have demonstrated that the sub-GeV dark matter with dark dipole interactions can be still safe from the direct-detection constraints, and the future low-threshold semiconductor experiments may play a significant role in constraining the dark dipole interactions.
}

\begin{document}

\maketitle

\section{Introduction}
The existence of dark matter (DM) has been well established from various cosmological and astrophysical observations. 
The particle nature of DM remains unknown despite great efforts of the observations, in particular its interactions to ordinary matter. 
The direct detection of nuclear recoils induced by DM scattering has severely constrained the elastic scattering cross section with nucleons. 
In particular, in the case of weakly interacting massive particle (WIMP) DM, direct detection experiments with liquid xenon (LXe) detectors have put the strongest constraint on the scattering cross section, which is tightly connected to the annihilation cross section determining the thermal relic abundance of DM. 
The scattering cross section should be smaller than about $10^{-47} \, \mathrm{cm}^2$ for the DM mass of $\sim 30 \, \mathrm{GeV}$~\cite{XENON:2023cxc,PandaX:2024qfu,LZ:2024zvo,XENON:2025vwd}.
LXe detector experiments have recently faced background events arising from neutrinos, often referred to as the neutrino fog.
Indeed, the direct measurement of nuclear recoils from solar neutrinos has been reported from the LXe detector experiment~\cite{XENON:2024ijk}.

Alternative DM models that are consistent with the current constraints from direct and indirect detection experiments have been proposed, and novel methods to detect DM interactions to ordinary matter have also been proposed for such alternative DM models in recent years. 
The constraints from the LXe detectors get weaker as the DM mass is below $\mathcal{O}(1) \, \mathrm{GeV}$ since the nuclear recoil energy is below a threshold energy. 
Thus, one of alternatives is the DM particle with the mass of (sub-)GeV. 
The nuclear recoils via the scattering with (sub-)GeV DM particles can be probed by similar experiments with different materials (such as argon detectors used in DarkSide~\cite{DarkSide:2018bpj}) and experiments with solid-state detectors (such as CRESST-III~\cite{CRESST:2019jnq}, Super-CDMS~\cite{SuperCDMS:2016wui,SuperCDMS:2020ymb}, and EDELWEISS~\cite{EDELWEISS:2019vjv,EDELWEISS:2020fxc}).
There are also alternative approaches for exploring the (sub-)GeV DM: the S2-only analysis for the nuclear recoils~(see \cite{XENON10:2011prx,XENON:2016jmt,XENON:2019gfn,XENON:2024znc} for XENON), the nuclear recoils with a Migdal electron~\cite{Ibe:2017yqa} and a Bremsstrahlung process~\cite{Kouvaris:2016afs}, and the DM-electron recoils~\cite{Essig:2011nj,Essig:2012yx,Essig:2017kqs}.

The phenomenology of the (sub-)GeV DM has been extensively studied: many of them are based on simplified models, in which a DM particle and a mediator particle are introduced.
A dark photon, which is a massive gauge boson of $U(1)_D$ kinematically mixing with the photon, is frequently introduced as the mediator particle.
If the DM particle is charged under $U(1)_D$, the annihilation process of DM particles into dark photons can determine the DM relic abundance. 
We can investigate the DM phenomenology through the dark photon: e.g., scattering with ordinary matter and annihilation into the SM particles.
Even if the DM particle is neutral under $U(1)_D$, it can couple to dark photon through dipole interactions.
Several studies have investigated the possibility that the DM particle possesses an electromagnetic dipole moment rather than a $U(1)_D$ dipole moment, called dipolar DM, originally proposed in Ref.~\cite{Sigurdson:2004zp} (and see Refs.~\cite{Pospelov:2000bq,Masso:2009mu,Banks:2010eh,Barger:2010gv,DelNobile:2014eta,Kadota:2014mea,Gondolo:2016mrz,Ibarra:2022nzm,Bose:2023yll,Ibarra:2024mpq,Eberl:2025kfm,Hochberg:2025rjs,Biswas:2026ahs} for latest studies on ``electromagnetic'' multipoles of fermionic DM particles for example). 
In this study, we consider the phenomenology of a fermionic DM with a $U(1)_D$ dipole moment.
This is an alternative to a simplified dark sector model, where the fermionic DM couples to the mediator dark photon in a standard way. 
In addition, dark sector models with confining gauge dynamics can naturally provide a fermionic DM that is neutral under $U(1)_D$~\cite{An:2009vq,Farina:2016ndq,Lonsdale:2017mzg,Lonsdale:2018xwd,Ibe:2018juk,Beauchesne:2020mih,Kamada:2020buc,Kamada:2021cow,Bodas:2024idn}, which has a $U(1)_D$ dipole moment analogous to the neutron.

The outline of this paper is as follows.
We briefly discuss the setup in this study in Section~\ref{sec:DDM}, the DM interacts with ordinary matter only through dipole interactions with dark photon particles. 
We also discuss the cosmological constraints such as the relic abundance in the thermal DM framework, the cosmological impacts by the late-time annihilation via the dark dipole interactions, and cosmic-ray observations.
In Section~\ref{sec:DD}, we discuss the direct detection experiments alternative to the conventional methods using the LXe detector.
We show the constraints on the dipole interactions from cosmology and the direct detection experiments in Section~\ref{sec:Results}.
Section~\ref{sec:Conc} is devoted to concluding this study.

\section{Dark Dipole Moment \label{sec:DDM}}

In this study, we consider the dark sector consisting of a Dirac fermion $\chi$ corresponding to the DM particle and dark photon $A'$. 
The dark photon is the massive gauge boson of a $U(1)_D$ symmetry and has a kinetic mixing with the hypercharge gauge boson. 
The dark photon may get massive by the Higgs mechanism or the St\"uckelberg mechanism.
The Lagrangian for dark photon is given by 
\begin{align}
    \mathcal{L} \supset 
    - \frac14 F'_{\mu \nu} F^{\prime \mu \nu} 
    + \frac{\epsilon}{2 \cos \theta_W} F'_{\mu \nu} B^{\mu \nu} 
    + \frac{1}{2} m_{A'}^2 A'^2_\mu \,, 
\end{align}
where $A'_\mu$ and $F'_{\mu \nu}$ denote the field and the field strength tensor of $U(1)_D$, respectively, and $B_{\mu \nu}$ is the hypercharge field strength tensor.
$\epsilon$ denotes the kinetic mixing parameter between the dark photon and the SM photon, $\theta_W$ is the Weinberg angle, and $m_{A'}$ denotes the mass of the dark photon.
The DM particle is assumed to be neutral under $U(1)_D$, and thus it does not couple to the dark photon through a vector current. 
Meanwhile, the DM particle can couple to the dark photon through the dipole interaction even if it is neutral under $U(1)_D$.
The dipole interactions of the DM are written as 
\begin{align}
    \mathcal{L} \supset 
    \frac{i}{2} d_{\chi} \overline{\chi} \sigma^{\mu \nu} \gamma^{5} \chi F'_{\mu \nu} 
    + \frac{1}{2} \mu_{\chi} \overline{\chi} \sigma^{\mu \nu} \chi F'_{\mu \nu}\,,
\end{align}
where $\chi$ denotes a four-component Dirac fermion, and $\sigma^{\mu \nu} = \frac{i}{2} [ \gamma^\mu, \gamma^\nu]$.
We assume that charge conjugation and parity transformation of the dark photon field are same as those of the SM photon field: we assume $F'_{\mu \nu}$ is a $CP$-even operator.
Under the assumption, the first term is the electric dipole interaction of $U(1)_D$, while the second term is the magnetic dipole interaction. 
$d_{\chi}$ and $\mu_{\chi}$ are the electric and magnetic moments of inverse mass dimension, respectively.

The DM particles couple to the dark photon through the dipole interactions and the dark photon mixes with the hypercharge gauge boson via the kinetic mixing. 
The DM particles couple to the SM fermions through the dark photon portal and the dipole interactions, as with the standard dark photon portal.
Thus, we can investigate this model through various processes similarly to the conventional dark photon portal, such as annihilation into the SM particles and scattering with the SM particles.
In this section, we discuss the existing constraints on this model from the annihilation processes.

\subsection{Relic Abundance}

The relic abundance is one of the key features of DM. 
We consider two scenarios for explaining the DM relic abundance: one is a thermal DM scenario, and the other is an asymmetric DM scenario  (see Refs.~\cite{Nussinov:1985xr, Barr:1990ca, Barr:1991qn, Kaplan:1991ah, Dodelson:1991iv, Kuzmin:1996he, Fujii:2002aj, Foot:2003jt,Foot:2004pq, Kitano:2004sv, Farrar:2005zd, Gudnason:2006ug, Kitano:2008tk, Kaplan:2009ag} for early work and also Refs.~\cite{Davoudiasl:2012uw, Petraki:2013wwa, Zurek:2013wia} for reviews).

When we assume the relic abundance of the DM particle is given by the freeze-out mechanism, it is determined by the annihilation process via the dipole interactions.
The relic abundance of the DM particle in the thermal DM scenario is approximately given by
\begin{align}
    \Omega_\chi h^2 = \frac{m_\chi s_0 Y_\infty}{ \rho_\mathrm{crit}/h^2}
    \simeq 8.63 \times 10^{-11} \,\mathrm{GeV}^{-2} \frac{x_f}{\sqrt{g_\ast} (a + 3b/x_f)} \,.
\end{align}
Here, $x_f = m_\chi/T_f$ is the ratio of the DM mass $m_\chi$ and the freeze-out temperature $T_f$, $g_\ast$ denotes the effective number of relativistic degrees of freedom used in the Boltzmann equation for $Y_\mathrm{DM} = n_\mathrm{DM}/s$ at freeze-out temperature.
We use $\rho_\mathrm{crit} \equiv 3 H_0^2 M_\mathrm{Pl}^2 \simeq 1.0539\times 10^{-5} h^2 \, \mathrm{GeV} \mathrm{cm}^{-3}$ and $s_0 \simeq 2891.2 \mathrm{cm}^{-3}$.
The velocity expansion of the annihilation cross section is $\sigma v = a + b v^2 + \cdots$, and hence $a$ and $b$ include the dark magnetic dipole moment and the dark electric dipole moment, respectively. 
We use the observed value of $\Omega_\chi h^2 = 0.12$ to constrain the dipole moments $\mu_\chi$ and $d_\chi$ when we consider the thermal DM scenario. 

There are two possible channels for the annihilation process via the dipole interactions: one is the annihilation into two dark photons, $\chi \overline \chi \to A' A'$, and the other is the annihilation into the SM fermions, $\chi \overline \chi \to f \overline f$.
In this study, we consider the DM particles with (sub-)GeV mass, and hence the annihilation channels into the SM bosons, such as $W$ bosons, are irrelevant. 
When the DM mass is larger than the dark photon mass, the annihilation into dark photons can determine the relic abundance of the DM particles. 
\begin{align}
    \sigma_{\chi \overline \chi \to A' A'} v 
    & = \frac{m_\chi^2}{4 \pi} \frac{\sqrt{1-r_{A'}}}{1-r_{A'}/2} \left[ \mu_\chi^4 \left( 1- \frac34 r_{A'} - \frac16 r_{A'}^2 \right) \right. \nonumber \\
    & \qquad \left. + d_\chi^4 (1-r_{A'}) (1-r_{A'}/2)^2 + 2 \mu_\chi^2 d_\chi^2 \left( 1 - r_{A'} + r_{A'}^2 + \frac14 r_{A'}^3 \right) \right]\,,
\end{align}
where $r_{A'} \equiv m_{A'}^2/m_\chi^2$.
Once the dark photon mass is negligible, the relic abundance by the freeze-out of the  annihilation $\chi \overline \chi \to A' A'$ is approximately given by
\begin{align}
    \Omega_\chi h^2 \simeq 0.01 \left( \frac{1 \, \mathrm{GeV}}{m_\chi} \right)^2 \left( \frac{10^{-15} \, e\,\mathrm{cm}}{\sqrt{\mu_\chi^2 + d_\chi^2}} \right)^4 \left( \frac{10}{g_\ast} \right)^{1/2} \,.
\end{align}
The annihilation process $\chi \overline \chi \to A' A'$ does not depend on the kinetic mixing parameter, and hence this process can control the relic abundance as far as $\epsilon \ll m_\chi \sqrt{\mu_\chi^2 + d_\chi^2}$.

Meanwhile, the relic abundance is determined by the annihilation process into the SM particles when the DM mass is smaller than the dark photon mass.
The annihilation cross section into the SM fermions in the non-relativistic limit is given by
\begin{align}
    \sigma_{\chi \overline \chi \to f \overline f} v 
    = \left(\frac{d^2_\chi v^2}{12} + \mu^2_\chi \right) \frac{\alpha \epsilon^2 \sqrt{1-r_f} \left(1+r_f/2 \right)}{\left(1-\frac14 r_{A'} \right)^2}\,,
\end{align}
where $v$ denotes the relative velocity of the DM particles, $\alpha$ denotes the fine structure constant, and $r_f \equiv m_f^2/m_\chi^2$.
The annihilation process through the dark magnetic dipole is $s$-wave, while the annihilation process through the dark electric dipole is $p$-wave.
Meanwhile, the annihilation cross section into hadrons, $\chi \overline \chi \to \text{hadrons}$, is also taken into account when the center-of-mass energy is above the threshold, and is given by 
\begin{align}
    \sigma_{\chi \overline{\chi} \to \mathrm{had}} 
    = \sigma_{\chi \overline{\chi} \to \mu \bar{\mu}} R (s = 4 m_\chi^2) \,.
\end{align}
Here, $\sigma_{\chi \overline{\chi} \to \mu \bar{\mu}}$ is the annihilation cross section into a pair of muons, and $R (s = 4 m_\chi^2)$ denotes the observed R-ratio.
We use the measured value of the R-ratio taken from Particle Data Group~\cite{ParticleDataGroup:2024cfk}, and $E_\mathrm{had} \simeq 0.15 \, \mathrm{GeV}$ is required as the threshold energy for annihilating to hadrons.
The explicit form of $a$ and $b$ for the annihilation process into fermions $f$ is given by 
\begin{align}
    a^{f}= \mu_{\chi}^2 \alpha \epsilon^2 \sqrt{1-\frac{m_f^2}{m_{\chi}^2}}\left(1+\frac{m_f^2}{2m_{\chi}^2}\right)\left(1-\frac{m_{A'}^2}{4m_{f}^2}\right)^{-2} \,, 
    \label{eq:a_value_ini} \\
    b^{f} = \frac{d_{\chi}^2}{12} \alpha \epsilon^2 \sqrt{1-\frac{m_f^2}{m_{\chi}^2}}\left(1+\frac{m_f^2}{2m_{\chi}^2}\right)\left(1-\frac{m_{A'}^2}{4m_{f}^2}\right)^{-2} \,.
    \label{eq:b_value_ini}
\end{align}

One may consider an asymmetric DM scenario for the case of $m_{A'} \lesssim m_\chi$.
The DM particles are strongly depleted into lighter dark-sector particles in order to reduce the symmetric part of the DM abundance, and dark photon plays a role of releasing entropy in the dark sector after the strong depletion of the DM particles through the kinetic mixing. 
Therefore, the mass spectrum $m_{A'} \lesssim m_\chi$ is required for consistent cosmology in the asymmetric DM scenario.
We do not specify how the particle--antiparticle asymmetry of DM particles is generated and related to that of baryons in this work (see Refs.~\cite{Shelton:2010ta,Falkowski:2011xh,Davoudiasl:2012uw,Bai:2013xga,Falkowski:2017uya,Ibe:2018juk,Hall:2019rld,Hall:2019ank,Kuwahara:2025eeo} for example). 

\subsection{Late-time Annihilation}

The annihilation of DM particles releases their energy into the electromagnetic plasma even in the late-time Universe.
The late-time energy injection from DM particles has been strongly constrained by the cosmological observations, such as big-bang nucleosynthesis (BBN) and cosmic microwave background (CMB).
In addition, the late-time annihilation produces cosmic-rays such as electron/positron flux and gamma-ray flux, which can be tested by cosmic-ray observations. 

As discussed in the previous subsection, there are two contributions from the dipole interactions to the annihilation processes: one is the $s$-wave from the dark magnetic dipole interaction, the other is the $p$-wave from the dark electric dipole interaction. 
The constraints from cosmology and cosmic-ray observations on the $p$-wave annihilation are much weaker than those in the $s$-wave annihilation. 
This is because the typical velocity of the DM particles is quite slow in the BBN epoch and later, and the $p$-wave annihilation cross section is suppressed by the DM velocity $v^2$.
We take into account the cosmological constraints on the dipole interactions from the CMB and BBN (see \cite{Henning:2012rm} for $s$-wave, and \cite{Diamanti:2013bia,Depta:2019lbe} for $p$-wave), and the cosmic-ray constraints on them (see \cite{Boudaud:2016mos} for $s$-wave, and \cite{Boudaud:2018oya} for $p$-wave).
Regarding the cosmic-ray observations, we use the constraints on the annihilation cross section of the DM particles into an electron-positron pair with assuming the propagation model with negligible reacceleration~\cite{Boudaud:2016mos,Boudaud:2018oya} and the cuspy DM profile, known as Navarro-Frenk-White (NFW)~\cite{Navarro:1996gj,Navarro:2003ew}.

The same amount of the DM antiparticle remains in the current Universe in the thermal DM scenario.
Meanwhile, in the asymmetric DM scenario, most of the DM antiparticle is assumed to be depleted, and the annihilation of the DM particles is quite a rare process in the late-time of the Universe. 
We simply ignore the constraints from cosmological observation and cosmic-ray observations for the asymmetric DM scenarios.

\subsection{Bounds on Dark Photon \label{sec:darkphoton}}

We introduce a massive dark photon as the mediator between the SM and dark sectors; there are several constraints on the dark photon parameters $\epsilon$ and $m_{A}^2$. 
When the dark photon is heavier than DM, the dark photon predominantly decays into the DM particles.
Since the dark photon has a kinetic mixing with the SM photon, dark photons can be produced at the terrestrial experiments, such as collider experiments and fixed-target experiments. 
The dark photon easily decays into a pair of DM particles, and hence the dark photon may leave missing energy signals, which can be explored at the LDMX experiment~\cite{Izaguirre:2017bqb,LDMX:2018cma}, and so on.

Meanwhile, when the dark photon is the lightest particle in the dark sector, the dark photon should decay into a pair of the SM particles, such as an electron-positron pair.
Depending on lifetime of dark photons, there are several ways to explore the dark photon signals.
When the (proper) lifetime is so short that dark photon decays inside the detector, the collider experiments (such as BaBar~\cite{BaBar:2014zli}, KLOE~\cite{KLOE-2:2011hhj,KLOE-2:2012lii,Anastasi:2015qla,KLOE-2:2016ydq}, and so on) have already put a stringent constraint. 
For the last decade, a new type of experiments has gained attention, so-called long-lived particle searches. 
When the (proper) lifetime lies in the range of $\mathrm{ns}$--$\mu\mathrm{s}$, the far-detector experiments (for example, FASER~\cite{Feng:2017uoz,FASER:2018eoc}, SHiP~\cite{SHiP:2015vad,Alekhin:2015byh}, DarkQuest~\cite{Gardner:2015wea,Berlin:2018pwi}, and so on) have a potential to explore the decay of dark photon with mass in the (sub-)GeV range.

Astrophysical observations have a chance to be probes for the dark photon for a longer lifetime: dark photon can easily escape from supernovae, which leads to an additional cooling process \cite{Chang:2016ntp}.
Besides the constraints from CMB and BBN on the annihilation cross section, an additional cosmological bound arises from the effective number of neutrino species $N_\mathrm{eff}$~\cite{Ibe:2019gpv}.
If dark photon remains in thermal contact with the electron-photon plasma after neutrino decoupling, its decay reheats only the electron-photon plasma and releases entropy only into the electromagnetic sector, leading to smaller $N_\mathrm{eff}$. 
The Planck measurement has precisely measured $N_\mathrm{eff}$~\cite{Planck:2018vyg}, and thus modification of $N_\mathrm{eff}$ from the SM value is strongly constrained. 

\section{Direct Detection Constraints\label{sec:DD}}

We consider the direct detection constraints on the DM particles interacting through dark dipole interactions. 
DM particles dispersed in the galaxy are non-relativistic, and the velocity distribution can be described by the Maxwellian distribution with typical velocity of $\mathcal{O}(10^{-3}) c$.
When the DM particles scatter with a nucleus in the detector at the direct detection experiments, the maximum momentum transfer is $q_\mathrm{max} = 2 \mu_{\chi N} v$, where $\mu_{\chi N}$ is the reduced mass of the DM and the nucleus. 
The elastic scattering of DM particle deposits small amounts of energy into a recoiling nucleus, the maximum recoil energy of nucleus is of $E_\mathrm{max} = 2 \mu_{\chi N}^2 v^2/m_N$. 
The conventional detectors utilizing the nuclear recoil (with the conventional analysis), such as the XENON experiment~\cite{XENON:2025vwd} and LUX-ZEPLIN (LZ) experiment~\cite{LZ:2024zvo}, have a lower energy threshold of $\mathcal{O}(1) \, \mathrm{keV}$.
Hence, such experiments lose their sensitivity to the DM particle with mass below $\mathcal{O}(10\,\mathrm{GeV})$.

In this section, we discuss alternative methods to detect (sub-)GeV DM particles, and we apply constraints and future sensitivities to the DM particles with dark dipole interactions. 
To begin with, we summarize relevant quantities for the DM-matter scattering process, such as the scattering cross section. 
Then, we discuss (i) the alternative processes in the DM-nucleus scattering which can be detected such as the Migdal effect, (ii) the DM-electron scattering, and (iii) the semiconductor detector. 

The scattering cross section of a DM particle is the most important quantity of the direct detection measurements. 
We now discuss the scattering process via the dipole interaction, although a dedicated treatment is required for each direct detection experiment.
The scattering cross section with a fermion $f$ via the dark dipole interaction is given by
\begin{align}
 	\sigma_f^\mathrm{EDM} &
    = \frac{\mu_{\chi f}^2}{16 \pi m_\chi^2 m_f^2} 64 \pi d_{\chi}^2 Q_f^2 \alpha \epsilon^2 m_{\chi}^2 \frac{m_f^2 q^2}{(q^2 +m_{A'}^2)^2}\,,
    \label{eq:EDM_cross_section} \\
 	\sigma_f^\mathrm{MDM} & 
    = \frac{\mu_{\chi f}^2}{16 \pi m_\chi^2 m_f^2} 16 \pi \mu_{\chi}^2 Q_f^2 \alpha \epsilon^2 m_{\chi}^2 \left( 1 - \frac{2 m_f}{m_\chi} \right) \frac{q^4}{(q^2 +m_{A'}^2)^2}\, . 
    \label{eq:MDM_cross_section}
\end{align}
Here, $\mu_{\chi f}$ denotes the reduced mass of the DM particle and the fermion $f$, $Q_f$ is the electromagnetic charge of the fermion, and $q$ is the momentum transfer.
$q^2$-dependence of the scattering cross section arises from how $\chi$ and $f$ couple to electric and magnetic fields.
The DM $\chi$ has only dipole interactions, while the fermion $f$ has a current interaction with the SM photon: the fermion $f$ couples to the electric field via charge while to the magnetic field via magnetic moment. 
This is the reason why the cross section $\sigma_f^\mathrm{EDM}$ via dark electric dipole is proportional to $q^2$, while $\sigma_f^\mathrm{MDM}$ via dark magnetic dipole is proportional to $q^4$.

Once we consider specific ultraviolet (UV) completion of the dark dipole interactions, there may be additional contributions to the scattering cross section.
For example, there might be a charge-radius operator of the DM particle, which is a dimension-six operator, which gives different $q^2$-dependence of the scattering cross section from the dipole interactions~\cite{DelNobile:2014eta}.
In addition, we consider the spontaneous breaking of $U(1)_D$ by the Higgs mechanism, there would be the Higgs--dark Higgs mixing leading to the Higgs portal interaction. 
Even though there is no direct coupling between the DM particle and the dark Higgs thanks to the $U(1)_D$ symmetry, there may be the Yukawa interaction among the DM particle, the dark Higgs, and an additional fermion charged under $U(1)_D$.
The Higgs portal interaction and the Yukawa interaction can also provide the scattering cross section between the DM particle and the nucleus via the loop of the dark Higgs and the new fermion, which is not suppressed by the momentum transfer $q$. 
These are model-dependent contributions, and we do not take into account such contributions in this study.

\subsection{Migdal Effect}
\label{sec:Migdal_effect}

Conventional direct detection experiments with a nuclear recoil assumes that atomic electrons around the recoiling nucleus immediately follow the motion of the nucleus. 
Once the nucleus recoils in a timescale shorter than the timescale of electron response, the electron cloud lags behind and experiences a sudden boost. 
This non-adiabatic behavior leads to atomic ionization or excitation, known as the Migdal effect, which is recently found in the neutron--nucleus scattering~\cite{Yi:2026fmf}. 
The final state ionization and excitation may enhance the detectability of (sub-)GeV DM detection experiments. 
Following the approach treating the nucleus and electron cloud coherently in Ref.~\cite{Ibe:2017yqa}, we compute the rate of the Migdal effect induced by DM scattering via the dark-dipole interactions.

When a DM particle elastically scatters off a nucleus of mass $m_T$, the nucleus receives momentum transfer $q_T=\sqrt{2m_T E_R}$ with the nuclear recoil energy $E_R$. 
The sudden nuclear recoil induces a boost on the electron with momentum $q_e \simeq m_e q_T / m_T$ and kinetic energy $E_e = q_e^2/(2m_e)$, which determine the probability of electronic excitation or ionization. 
The scattering cross section is decomposed into two parts: one is the nuclear recoil, and the other is the transition of the electron cloud. 
The transition of the electron cloud is encoded in a factor $Z_{FI}(q_e)$, and the electron cloud is labelled by the principal number $n$ and the orbital angular momentum $\ell$,
\begin{equation}
	\sum_{F}|Z_{FI}|^2 = |Z_{II}|^2 + 	
	\sum_{n,\ell,n',\ell'}p^d_{q_e}(n \ell \to n' \ell') + 
	\sum_{n,\ell}\int \frac{\mathrm{d} E_e}{2\pi} \frac{\mathrm{d}}{\mathrm{d} E_e} p^c_{q_e}(n \ell \to E_e) \,,
	\label{eq:Migdal_transition}
\end{equation}
where $I$ and $F$ represent the initial and final electronic states.
The probability for the electrons unaffected is denoted by $|Z_{II}|^2$, namely $|Z_{II}|^2 \simeq 1 + \mathcal{O}(q_e^2 \langle r^2 \rangle)$. 
Meanwhile, $p^d_{q_e}$ describes bound-state excitations, and $p^c_{q_e}$ describes ionization into scattering states.
We discuss the ionization form factor in detail in Appendix~\ref{app:ionization}.

The differential cross section for DM-nucleus scattering accompanied by a Migdal transition is obtained by multiplying the nuclear-recoil cross section with the electronic transition:
\begin{equation}
    \label{eq:Migdal_corss_section}
    \frac{\mathrm{d} \sigma}{\mathrm{d} E_R} = \frac{\mathrm{d} \sigma_0}{\mathrm{d} E_R}\sum_{F}|Z_{FI}|^2,
\end{equation}
where $\mathrm{d} \sigma_0 / \mathrm{d} E_R$ is the nuclear-recoil differential cross section. 
For the dipole interactions considered in this work, its explicit form is~\cite{DelNobile:2021wmp}
\begin{align}
	\frac{\mathrm{d} \sigma_0}{\mathrm{d} E_R} &= \frac{\epsilon^2 \alpha}{2} \left( \frac{q_T^2}{q_T^2 + m^2_{A'}}\right)^2 \frac{m_{T}}{m^2_{N}} \frac{1}{v^2} \left\{ \mu^2_{\chi}\left[\left( \frac{1}{m^2_\chi} - \frac{1}{\mu^2_{T\chi}}+\frac{4 v^2}{q_T^2}\right)m^2_N F^{(p,p)}_M(q_{T}^2) + 4 F^{(p,p)}_\Delta(q_{T}^2) \right. \right. \nonumber \\ 
	& \qquad \left. \left. + \frac{1}{4}\sum_{N,N'=p,n}g_N g_{N'}F^{(N, N')}_{\Sigma'}(q_{T}^2) - \sum_{N=p,n}2 g_N F^{(N,p)}_{\Sigma', \Delta}(q_{T}^2)\right] + 4 d^2_{\chi} \frac{m^2_N}{q_T^2} F^{(p,p)}_M(q_{T}^2) \right\} \,, \label{eq:diff_cross_section_nucleus}
\end{align}
with $\mu_{T\chi}=m_Tm_{\chi}/(m_T+m_{\chi})$ the reduced mass of DM and nucleus. $N=p,n$ correspond to the proton and neutron. 
$g_N$ is the Land\'e $g$-factor of proton and neutron. 
A detailed derivation of Eq.~\eqref{eq:diff_cross_section_nucleus} is shown in Appendix~\ref{app:DM-Nucleus}.
The nucleus response functions $F^{(p,p)}_M$, $F_{\Delta}^{(p,p)}$, $F_{\Sigma'}^{(N,N')}$ and $F_{\Sigma', \Delta}^{(N,p)}$~\cite{DelNobile:2021wmp} can be numerically calculated through Ref.~\cite{Anand:2013yka}. 
The differential scattering rate of the DM-nucleus scattering process is defined by
\begin{equation}
    \label{eq:Migdal_reaction_rate}
    \frac{\mathrm{d}R}{\mathrm{d}E_R \mathrm{d}v}=n_\mathrm{DM}\frac{\mathrm{d} \sigma}{\mathrm{d}E_R}vf(v)=n_\mathrm{DM}\frac{\mathrm{d} \sigma_0}{\mathrm{d}E_R}vf(v)\sum_{F}|Z_{FI}|^2 \equiv \frac{\mathrm{d}R_0}{\mathrm{d}E_{R}dv}\sum_{F}|Z_{FI}|^2 \,.
\end{equation}
Here, $n_\mathrm{DM} \equiv \rho_{\chi} / m_{\chi}$ is the DM number density, and $f(v)$ is the DM speed distribution function. 
For calculating the rate, we assume an isotropic Maxwell-Boltzmann distribution for the DM speed in the galactic rest frame. 
\begin{equation}
	f(v) = \frac{1}{K} \exp \left[ - \frac{|\vec{v}+\vec{v}_E|^2}{v_0^2}  \right] \Theta(v_\mathrm{esc} - |\vec{v}+\vec{v}_E|) \,.
\end{equation}
Here, we take the typical velocity $v_0 = 238 \, \mathrm{km}/\mathrm{s}$ and the escape velocity $v_\mathrm{esc} = 544 \, \mathrm{km}/\mathrm{s}$~\cite{Baxter:2021pqo}.
The average velocity of the Earth relative to the DM halo is computed as the sum of the velocity of the local standard of rest, the peculiar velocity of the Sun with respect to the DM halo, and the Earth's velocity with respect to the Sun; which is implemented in \texttt{wimprates}~\cite{wimprates:2023}.
$K$ denotes the normalization constant of the velocity distribution. 
We can define the differential scattering rates for ionization and excitation processes by combining Eqs.~\eqref{eq:Migdal_transition} and \eqref{eq:Migdal_reaction_rate}.
We find the electron energy spectrum for each contribution.
\begin{align}
	\label{eq:diff_migdal_ion}
	\frac{\mathrm{d} R}{\mathrm{d} E_R \mathrm{d} E_{e}  \mathrm{d} v} \bigg|_\mathrm{ion} & \simeq \frac{\mathrm{d} R_0}{\mathrm{d} E_R  \mathrm{d} v}\times \frac{1}{2\pi}\sum_{n,\ell}\frac{\mathrm{d}}{\mathrm{d} E_e}p^c_{q_e}(n \ell \to (E_e+E_\mathrm{dex}-E_{n \ell})) \,, \\
	\label{eq:diff_migdal_exc}
	\frac{\mathrm{d} R}{\mathrm{d} E_R \mathrm{d} E_{e}  \mathrm{d} v} \bigg|_\mathrm{exc} & \simeq \frac{\mathrm{d} R_0}{\mathrm{d} E_R \mathrm{d} v}\times \sum_{n,\ell, n', \ell'} p^d_{q_e}(n \ell \to n' \ell')\times \delta(E_{e}+E_\mathrm{dex} - \Delta E_{n \ell \to n' \ell'}) \,,
\end{align}
where $E_\mathrm{dex}$ is the de-excitation energy and $E_{n\ell}$ the bound-state energy. 
Again, we discuss the functions $p^{c}_{q_e}$ and $p^{d}_{q_e}$ in detail in Appendix~\ref{app:ionization}.

We obtain the total event rate for the Migdal effect by integrating the differential rates, Eqs.~\eqref{eq:diff_migdal_ion} and \eqref{eq:diff_migdal_exc}, over kinetic variables $E_R$, $E_e$, and $v$. 
We compare our predicted rate with the DarkSide measurement reported in Ref.~\cite{DarkSide:2022dhx}. 
This yields the corresponding constraints on the dipole couplings $\mu_{\chi}$ and $d_{\chi}$ from the Migdal effect.
The DarkSide experiment uses a liquid argon target in its detector to search for the nuclear recoils from the DM-nucleon elastic scattering, providing a powerful probe for constraining dark matter parameters. 
A recent analysis by DarkSide-50 has exclusively focused on the electron ionization signal~\cite{DarkSide:2022dhx} (corresponding to Eq.~\eqref{eq:diff_migdal_ion}), leading to a constraint on the spin-independent DM-nucleus cross section $\sigma_\mathrm{SI}$ for given DM mass. 
In our analysis, we compute the total event rate from the experimental upper bound on the scattering cross section $\sigma_\mathrm{SI}$ for given DM mass, denoted by $R_\mathrm{exp}$, and then we compare the total event rate $R_\mathrm{exp}$ with the total event rate induced by the DM dipole interactions, denoted by $R_\mathrm{th}$.
Once the event rate $R_\mathrm{th}$ exceeds $R_\mathrm{exp}$, we set the upper bounds on the DM dipole couplings, $\mu_{\chi}$ and $d_{\chi}$. 
We note that we obtain the constraints on the dark dipole moments by comparing the total event rate, rather than the differential event rate as a function of the electron recoil energy.
Therefore, our constraints would be conservative, and a more detailed analysis would improve the bounds on the dark dipole interactions. 
We also incorporate the constraints from conventional nuclear-recoil searches at DarkSide-50 for DM masses above $0.5$ GeV.

\subsection{Dark Matter--Electron Scattering}

As mentioned in the beginning of this section, a DM particle with sub-GeV mass does not transfer enough energy to produce nuclear recoils to be detected in LXe detectors. 
In contrast, the kinetic energy of the DM particle, $m_\chi v^2 /2 \simeq 50 \, \mathrm{eV} \times (m_\chi /100 \, \mathrm{MeV})$, is sufficient for inelastic atomic processes, such as ionization and excitation of atomic electrons. 
The resulting electron recoils and secondary photons provide sensitive probes of light DM. 

To compare with the conventional approach, we decompose the DM-electron scattering cross sections, given by Eqs.~\eqref{eq:EDM_cross_section} and \eqref{eq:MDM_cross_section} with electron, into a reference cross section $\overline{\sigma}_e$ and a momentum-dependent DM form factor $F_\mathrm{DM}(q)$. 
\begin{align}
	\label{eq:electron_sigma}
	\overline{\sigma}_e \equiv \sigma_e (q^2 = \alpha^2 m_e^2) \,, \qquad 
	|F_\mathrm{DM}(q)|^2 \equiv \frac{\sigma_e (q^2)}{\overline{\sigma}_e} \,.
\end{align}
A typical energy scale in atoms is given by the Bohr radius, $a_0^{-1} \sim \alpha m_e$, and hence the reference cross section $\overline{\sigma}_e$ characterizes a typical size of DM-electron scattering cross section. 
The momentum-transfer dependence is important for ionizing an electron from an atom: combining the form factor $F_\mathrm{DM}(q)$ and the transition of electron states in the atomic physics, we obtain the final cross section for the ionization.
Given scattering cross section in Eqs.~\eqref{eq:EDM_cross_section} and \eqref{eq:MDM_cross_section}, we find the form factors for 
\begin{align}
 	\left| F_\mathrm{DM}^\mathrm{EDM}(q) \right| =\left(\frac{q}{\alpha m_e}\right) \left(\frac{m_{A'}^2+m_e^2 \alpha^2}{m_{A'}^2+q^2}\right) \,, \quad 
 	\left| F_\mathrm{DM}^\mathrm{MDM}(q) \right| = \left(\frac{q}{\alpha m_e}\right)^2\left(\frac{m_{A'}^2+m_e^2 \alpha^2}{m_{A'}^2+q^2}\right) \,.
	\label{eq:form_factor_EDM}
 \end{align}
In DM-electron scattering with a vector mediator coupling to the SM fermion through a current interaction, the momentum-transfer dependence of the form factor is $F_\mathrm{DM} \sim 1$ for massive mediator and $F_\mathrm{DM} \sim \alpha^2 m_e^2 / q^2$ for light mediator. 
Meanwhile, the momentum-transfer dependence in dark dipole DM framework is different from the conventional vector-mediator models: $F^\mathrm{EDM}_\mathrm{DM}(q) \sim q/\alpha m_e$ and $F^\mathrm{MDM}_\mathrm{DM}(q) \sim q^2/\alpha^2 m_e^2$ for massive mediator $m_{A'} \gg q, \alpha m_e$, $F^\mathrm{EDM}_\mathrm{DM}(q)\sim \alpha m_e/q$ and $F^\mathrm{MDM}_\mathrm{DM}(q) \sim 1$ for for light mediator $m_{A'} \ll q, \alpha m_e$.

To compute the ionization probability, we now discuss the transition from a bound state $i$ to a scattering state with momentum $k$ (and energy $E_k = k^2/2m_e$).
To overcome the binding energy of an electron in an atom, the DM velocity must exceed the minimum value, which is determined by the energy conservation,
 \begin{equation}
	\label{eq:min_v_electron}
	v_\mathrm{min}=\frac{E^{i}_B+E_k}{q}+\frac{q}{2m_{\chi}} \,,
\end{equation}
where $E^{i}_{B}$ is the binding energy of the bound state $i$.
The velocity-averaged differential cross section for the DM-electron scattering with ionization is computed as follows~\cite{Essig:2011nj}.
\begin{equation}
	\label{eq:diff_electron}
	\frac{\mathrm{d}\langle \sigma^{i} v \rangle}{\mathrm{d} \ln E_k} = \frac{\overline{\sigma}_e}{8\mu^2_{\chi e}} \int \mathrm{d}q \, q |f^{i}_\mathrm{ion}(E_k,q)|^2 |F_\mathrm{DM}(q)|^2 \eta(v_\mathrm{min}) \,,
\end{equation}
where the mean inverse speed of DM is given by
\begin{equation}
	\label{eq:eta_definition}
	\eta(v_\mathrm{min}) = \int_{v>v_\mathrm{min}} \frac{f(v)}{v}\mathrm{d}^3 v \,,
\end{equation}
and $f(v)$ is the DM velocity distribution function, again. 
The cross section includes transitions from the bound state $i$ to scattering states, which are implemented as the electron ionization form factor $f^{i}_\mathrm{ion}(E_k,q)$.
\begin{align}
	\label{eq:ionization_form_factor_total}
	|f^{i}_\mathrm{ion}(E_k,q)|^2 & = \sum_{\mathrm{occupied}} \sum_{\ell,m}\frac{2 k^3}{(2\pi)^3}|f^{i\to k,\ell,m}(\mathbf{q})|^2 \,,  \\
	\label{eq:ionization_form_factor}
	|f^{i\to k,\ell ,m}(\mathbf{q})|^2 &= \bigg| \int \mathrm{d}^3 \mathbf{r} \, \psi_i(\mathbf{r}) 4 \pi j_{l} (k r) Y^{\ast}_{\ell m}(\theta_{\mathbf{r}},\phi_{\mathbf{r}}) e^{i \mathbf{q}\cdot \mathbf{r}} \bigg|^2 \,,
\end{align}
where $\ell$ and $m$ are the angular quantum numbers of the final electron state. 
The wavefunction of the final state (free) electron is written by using the spherical Bessel function of the first kind $j_\ell$ and the spherical harmonics $Y_{\ell m}$, while 
$\psi_i(\mathbf{r})$ denotes the wavefunction of the initial bound state.
Since we sum over the angular momentum variables of both initial and final states, the ionization form factor does not depend on the direction of $\mathbf{q}$.
We note that the ionization cross section is factorized into the part of the DM-electron scattering and the atomic reaction part: the atomic reaction is unique once the momentum transfer to the bound electron is given.
The total event rate per target mass $R_\mathrm{ion}/m_T$ for the ionization is given by 
\begin{equation}
	\label{eq:total_rate_electron}
	\frac{R_\mathrm{ion}}{m_T} = 
	\frac{n_\mathrm{DM}}{m_T}\frac{\overline{\sigma}_e}{8\mu^2_{\chi e}} \sum_{i} \int \mathrm{d} q \mathrm{d}E_k \, \frac{q}{E_k}|f^{i}_\mathrm{ion}(E_k,q)|^2 |F_\mathrm{DM}(q)|^2 \eta(v_\mathrm{min}) \,,
\end{equation}
where $m_T$ is the detector target mass. 
We use the numeric values of $|f^{i}_\mathrm{ion}(E_k,q)|^2$ implemented in public codes, \texttt{wimprates}~\cite{wimprates:2023} and \texttt{QEDark}~\cite{Essig:2015cda}.

In the following analysis, we consider direct-detection experiments using electron recoils to put constraints on the dark-dipole DM scenario: XENON10 and XENONnT. 
XENON10 data consists of events that have ``S2''-only (namely, without a prompt scintillation signal) signals with multiple electrons in the final state~\cite{Essig:2012yx,Essig:2017kqs}.
XENONnT has also reported new results on the search for DM-electron scattering using the data acquired in 2021-2022~\cite{XENON:2024znc}.
We use the conservative bound from the blinded search of the data acquired in September-October 2021, which corresponds to an exposure of $16.5$~days~\cite{XENON:2024znc}.
We compute the ionization rate from their bound on the DM-electron cross section, and we obtain the constraints on the dipole moments $\mu_\chi$ and $d_\chi$.
We will give the detail of the constraints from existing data of XENON10 and XENONnT in the next section.

\subsection{Semiconductor Detector}

We consider constraints from direct-detection experiments using semiconductor crystals as a detector in this subsection. 
Semiconductor materials suit for probing DM with a mass of MeV to sub-GeV: the energy gap (the threshold energy to excite electrons in the valence band into the conduction band) is of $\mathcal{O}(1 \mathrm{eV})$. 
This energy threshold is $\mathcal{O}(10)$ times lower than the ionization threshold energy in LXe, making sensitivity to DM masses down to the MeV scale. 
A key difference of the direct detections with semiconductor crystals from those with ionization is that the initial state of electron is written in terms of the Bloch wavefunction in a periodic lattice. 
The many-body nature of crystals modifies the electronic wavefunctions, leading to a crystal form-factor $f_\mathrm{crystal}$ to incorporate an effect of a band structure~\cite{Essig:2015cda}.

As with ionization, the detector with semiconductor crystals requires a kinetic condition for minimum velocity of DM, 
\begin{equation}
	v_\mathrm{min}=\frac{E_e}{q}+\frac{q}{2m_{\chi}} \,,
	\label{eq:min_v_semiconductor}
\end{equation}
in order to excite an electron in the valence band. 
Here, $E_e$ is the energy of an electron in the conduction band and $q$ denotes the momentum transfer. 
The kinetic condition is very close to that in the ionization process, and the electron-DM scattering does not change at microscopic level.
The difference from the ionization arises from an electron wavefunction, namely the crystal form factor denoted by $f_\mathrm{crystal}$. 

The differential excitation rate is computed by incorporating an electron wavefunction in crystal in a similar way to the differential ionization rate. 
\begin{equation}
	\label{eq:diff_semiconductor}
	\frac{\mathrm{d} R_\mathrm{crystal}}{\mathrm{d} \ln E_e} 
	= n_\mathrm{DM} \overline{\sigma}_e N_\mathrm{cell} \alpha \frac{m_e^2}{\mu_{\chi e}^2}\int \mathrm{d} \ln q \frac{E_e}{q} |f_\mathrm{crystal}(q, E_e)|^2 |F_\mathrm{DM}(q)|^2 \eta(v_\mathrm{min}) \,,
\end{equation}
where $f_\mathrm{crystal}(q,E_e)$ denotes the crystal form factor, and $N_\mathrm{cell} \equiv m_T / m_\mathrm{cell}$  is the number of unit cells in the crystal target with mass $m_{T}$. 
For silicon and germanium detectors, the cell masses $m_\mathrm{cell}$ are 52.33 GeV and 135.33 GeV, respectively~\cite{Essig:2015cda}.
The electron energy level in a crystal is characterized by the band index $i$ and a continuum wave vector $\mathbf{k}$ in the first Brillouin zone (BZ), and the electron wavefunction in a crystal is written in the Bloch form,
\begin{equation}
	\label{eq:electron_wavefunction_semiconductor}
	\psi_{i \mathbf{k}}(\mathbf{x}) = \frac{1}{\sqrt{V}} \sum_{\mathbf{G}} u_i(\mathbf{k}+\mathbf{G}) e^{i(\mathbf{k}+\mathbf{G})\cdot \mathbf{x}} \,,
\end{equation}
where $\mathbf{G}$ runs over the reciprocal lattice vectors and $V$ is the volume of the crystal target. 
The normalization of the wavefunction implies that the coefficients $u_i(\mathbf{k}+\mathbf{G})$ satisfy the following condition. 
\begin{equation}
	\label{eq:normalization}
	\sum_{\mathbf{G}}|u_i(\mathbf{k}+\mathbf{G})|^2 = 1 \,.
\end{equation}
The crystal form factor is obtained as follows~\cite{Essig:2015cda}.
\begin{align}
	\label{eq:crystal_form_factor}
	|f_\mathrm{crystal}(q,E_e)|^2&=\frac{2\pi^2(\alpha m_e^2 V_\mathrm{cell})^{-1}}{E_e}\sum_{i,i'}\int_\mathrm{BZ} \frac{V_\mathrm{cell}\mathrm{d}^3\mathbf{k}}{(2\pi)^3}  \frac{V_\mathrm{cell}\mathrm{d}^3\mathbf{k}'}{(2\pi)^3} \nonumber \\ 
	& \times E_e \delta(E_e-E_{i'\mathbf{k}'}+E_{i\mathbf{k}})\sum_{\mathbf{G}'}q\delta(q-|\mathbf{k}'-\mathbf{k}+\mathbf{G}'|)|f_{[i\mathbf{k},i'\mathbf{k}',\mathbf{G}']}|^2 \,, 
\end{align}
where $V_\mathrm{cell} \equiv V/N_\mathrm{cell}$ is the volume of the unit cell and $E_{i \mathbf{k}}$ is the energy of a level labelled by the wave vector $\mathbf{k}$ and the band index $i$. 
We define the form factor for excitation from a valence level $i, \mathbf{k}$ to a conduction level $i', \mathbf{k}'$ as follows. 
\begin{align}
	\label{eq:crystal_form_factor_component}
	f_{[i\mathbf{k},i'\mathbf{k}',\mathbf{G}']} & = \sum_{\mathbf{G}}u^{*}_{i'}(\mathbf{k}'+\mathbf{G}+\mathbf{G}')u_{i}(\mathbf{k}+\mathbf{G}) \,.
\end{align}
Information about the crystal target, such as the band structure, band dispersion, and overlaps of the Bloch states, is entirely encoded in the crystal form factor $f_\mathrm{crystal}(q,E_e)$.
Meanwhile, as mentioned, the microscopic DM interactions enter only through $\overline{\sigma}_e$ and $F_\mathrm{DM}(q)$.

Combining them, the total event rate per target mass $m_\mathrm{cell}$ is obtained as follows.
\begin{equation}
	\label{eq:total_rate_semiconductor}
	\frac{R_\mathrm{crystal}}{m_T} = \alpha \overline{\sigma}_e \frac{n_\mathrm{DM}}{m_\mathrm{cell}}  \frac{m_e^2}{\mu_{\chi e}^2}\int \mathrm{d} \ln E_e  \mathrm{d} \ln q \frac{E_e}{q} |f_\mathrm{crystal}(q, E_e)|^2 |F_\mathrm{DM}(q)|^2 \eta(v_\mathrm{min}) \,.
\end{equation}
In the numerical analysis, we consider the semiconductor detectors using silicon target and germanium target. 
We take the list of numerical values for the crystal form factors $|f_\mathrm{crystal}(q,E_e)|^2$ for the Silicon and Germanium detectors from~\cite{asingal14QCDark,Dreyer:2023ovn,Essig:2015cda}. 
With the event rate evaluated through Eq.~\eqref{eq:total_rate_semiconductor}, we compare our prediction to the expected sensitivity, $R_\mathrm{th}/m_T = 3.6\ \mathrm{events\cdot year^{-1}\cdot kg^{-1}}$~\cite{Essig:2015cda} and set the constraints on the dark dipole moments. 
For silicon-based detectors, we derive constraints on dark dipole interactions through the SENSEI experiment~\cite{SENSEI:2023zdf} and the DAMIC-M experiment~\cite{DAMIC-M:2025luv}, which are particularly sensitive to sub-GeV dark matter. 
Both SENSEI and DAMIC-M use the skipper-CCDs, which observe the recoiled electron in silicon~\cite{SENSEI:2023zdf}. 
Both experiments have reported the limits on the reference electron-scattering cross section $\overline{\sigma}_e$ for dark-matter form factors $F_\mathrm{DM} = (\alpha m_e / q)^2$ and $F_\mathrm{DM} = 1$. 
Using Eq.~\eqref{eq:total_rate_semiconductor}, we translate these bounds into constraints on the total event rate and compare them with our predicted rates to constrain $\mu_{\chi}$ and $d_{\chi}$. 
While for germanium targets, we instead compare our predicted event rate with the projected sensitivity $R_\mathrm{th}/m_T = 3.6\ \mathrm{events\cdot year^{-1}\cdot kg^{-1}}$~\cite{Essig:2015cda} and then set the constraints on the dark dipole moments. 

\section{Results \label{sec:Results}}
In this section, we present the constraints on dark dipole moments of DM obtained from direct-detection experiments and cosmological observations (including relic abundance for the case of thermal DM). 
This section is composed of two subsections: one is for the constraints on the dipole moments $d_{\chi}$ and $\mu_{\chi}$ for given kinetic mixing parameter $\epsilon = 10^{-3}$, the other is for the constraints on the mixing parameter $\epsilon$ under assuming that the DM particle has the nuclear magneton like dipole moments.
For both cases, we assume two patterns of mass hierarchy of the dark-sector particles:
\begin{itemize}
    \item The fermionic DM $\chi$ is lightest in the dark sector, $m_{A'}=3m_{\chi}$, and its abundance is explained by thermal freeze-out.
    \item The dark photon is lightest in the dark sector, $m_{A'}=0.1m_{\chi}$, and the DM abundance can be explained as the asymmetric DM.
\end{itemize}
The constraints from the direct-detection experiments and the cosmological observations are summarized in Figures~\ref{fig:constraints_dipole} and \ref{fig:epsilon_constraint}. 

\subsection{Constraints on the Dark Dipole Moments}

\begin{figure}[tp]
	\includegraphics[width=0.48\textwidth]{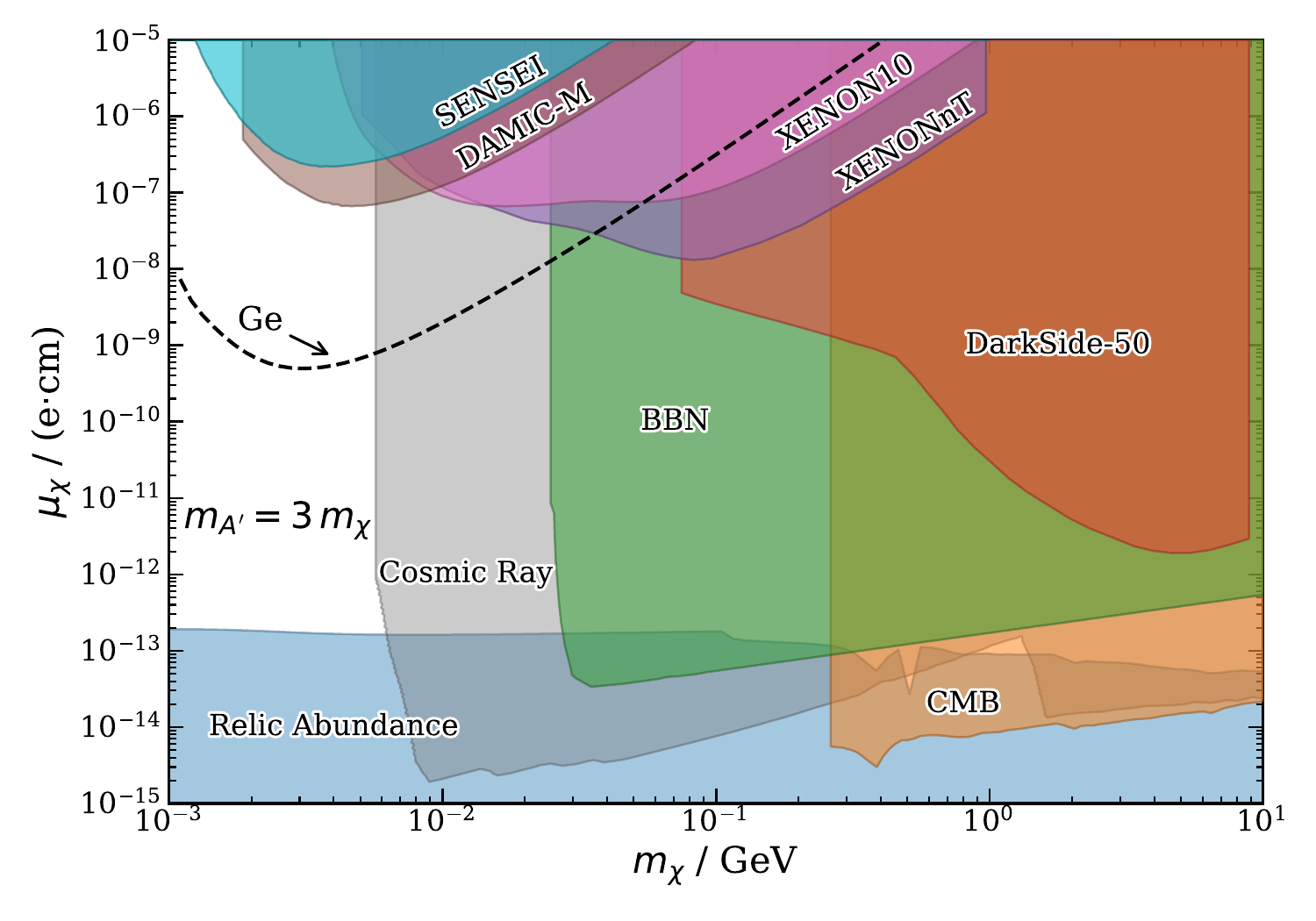}
    \includegraphics[width=0.48\textwidth]{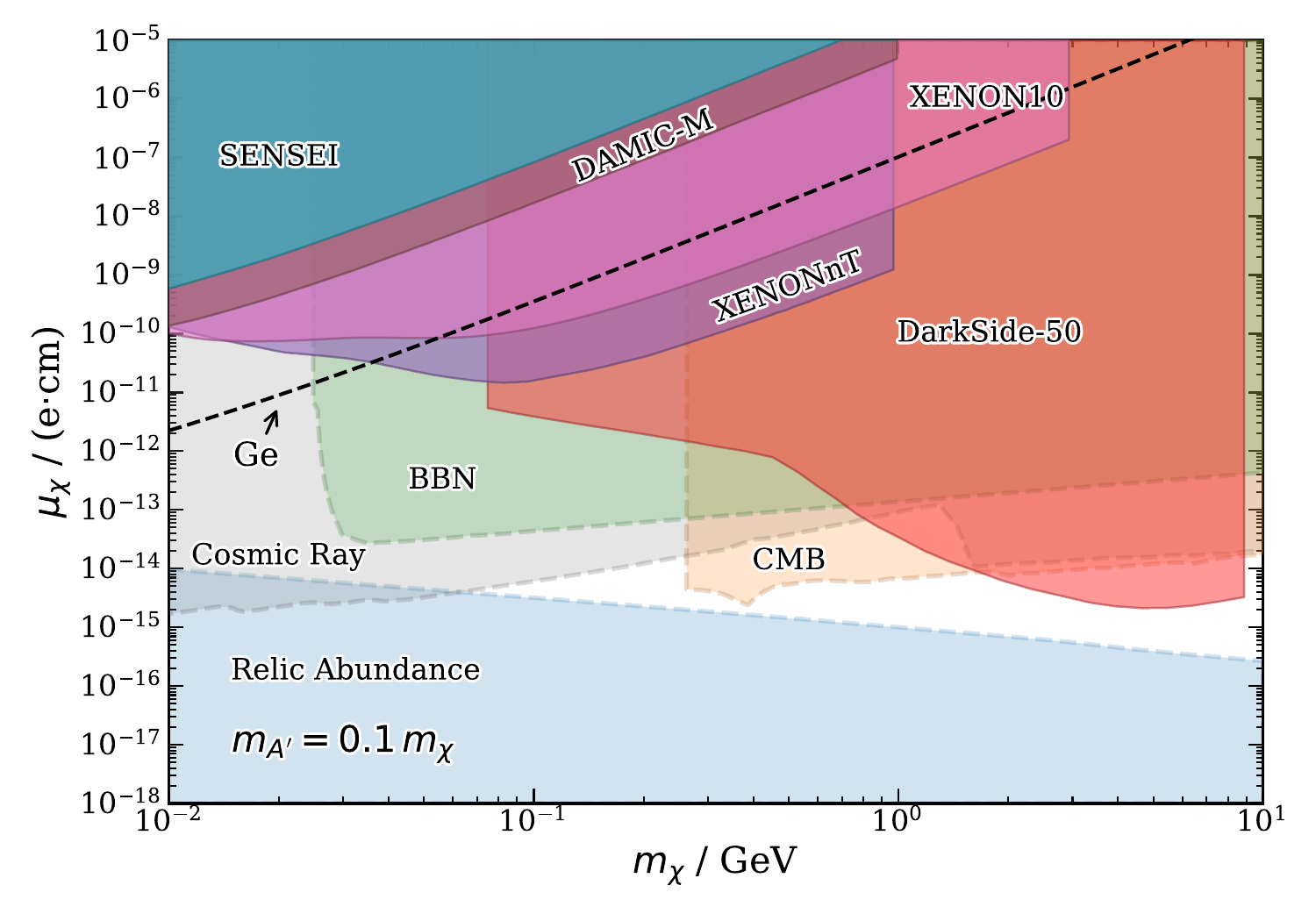} \\ \noindent
	\includegraphics[width=0.48\textwidth]{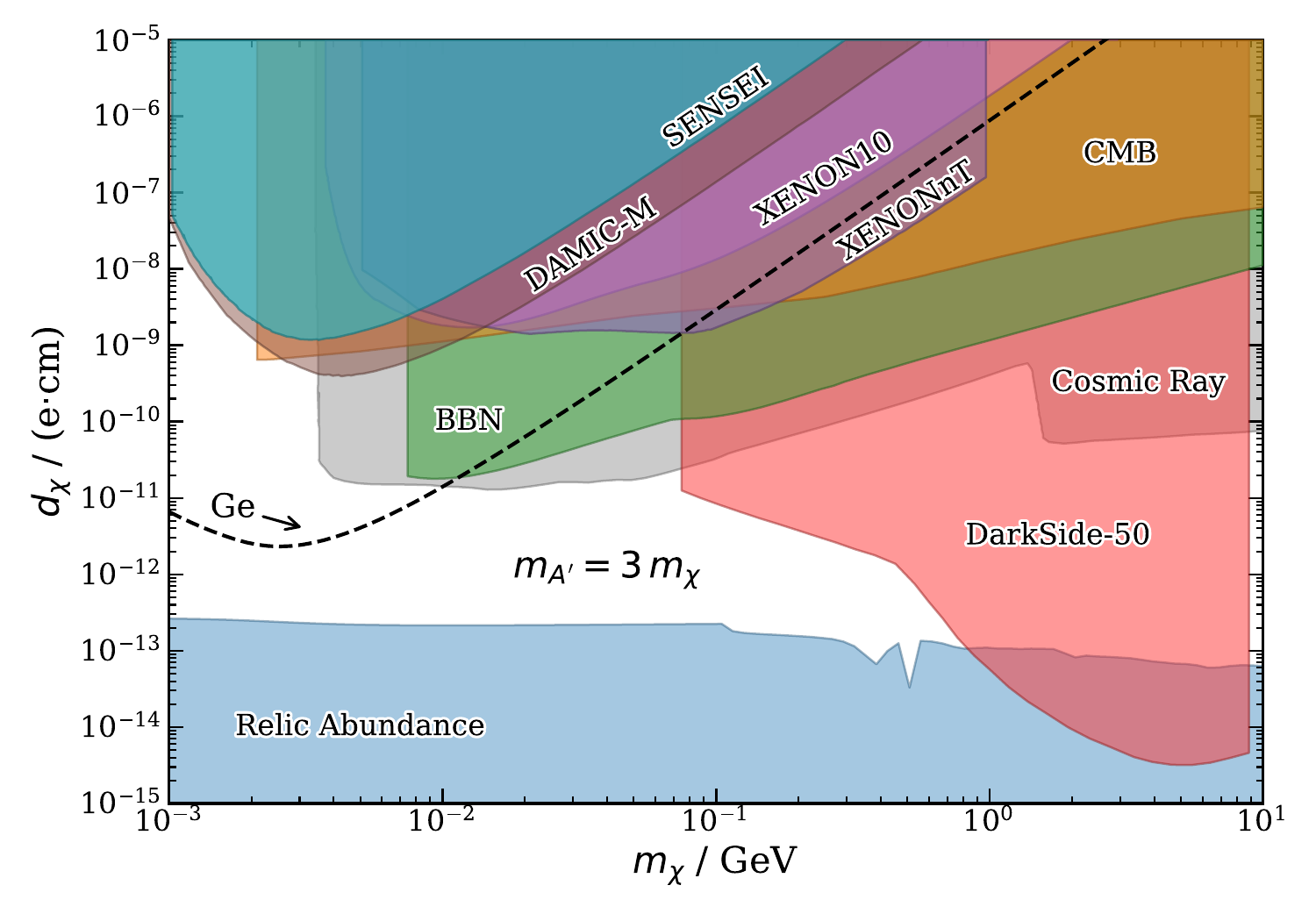}
    \includegraphics[width=0.48\textwidth]{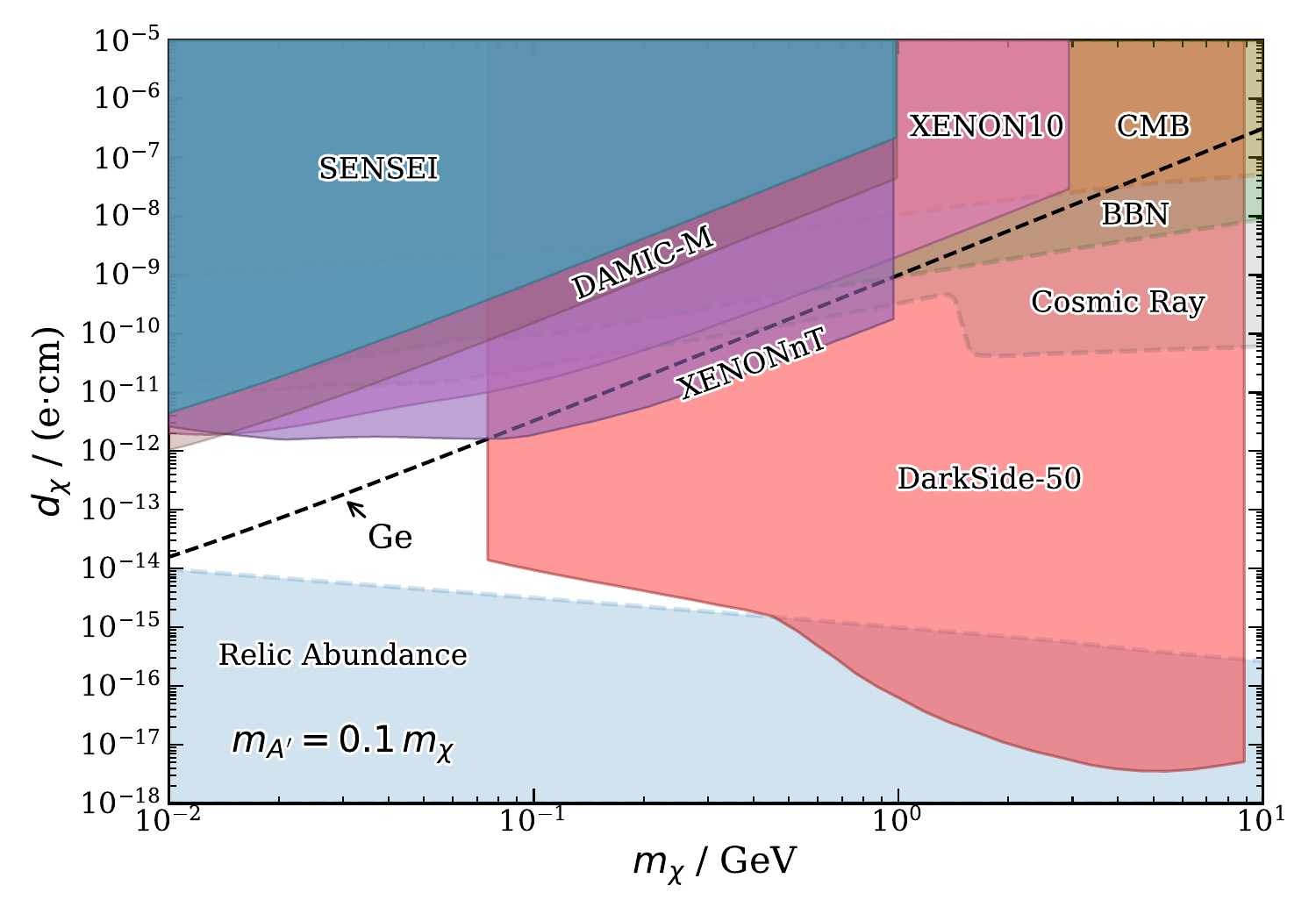}
	\caption{
    Constraints on the dark dipole moments with $\epsilon = 10^{-3}$: the electric dipole $d_{\chi}$ (top) and the magnetic dipole $\mu_{\chi}$ (bottom). 
    The left column corresponds to $m_{A'} = 3m_{\chi}$, where $\chi$ is the lightest dark-sector particle. 
    Meanwhile, the right column corresponds to $m_{A'} = 0.1m_{\chi}$, where $A'$ is the lightest dark-sector particle. 
    Constraints from the direct-detection experiments include the Migdal effect (DarkSide-50: bright red shaded), DM-electron scattering (XENON10: pink shaded, XENONnT: purple shaded), and semiconductor detector (SENSEI: cyan shaded, DAMIC-M: brown shaded, projected germanium detector: dashed line).
    Cosmological and astrophysical constraints include the relic abundance (steel blue shaded), the CMB (orange shaded), the BBN (green shaded), and cosmic-rays (gray shaded): these bounds can be easily avoided in the case of asymmetric DM, so we put them in relatively transparent colors and dashed boundaries in the right panels.
	}
	\label{fig:constraints_dipole}
\end{figure}

We begin with considering the constraints on the dark dipole moments. 
We fix the kinetic mixing parameter to $\epsilon = 10^{-3}$ in this subsection. 
As mentioned, we assume the ratio of the dark photon mass $m_{A'}$ and the DM mass $m_\chi$, $m_{A'}=3m_{\chi}$ and $m_{A'}=0.1m_{\chi}$. 
Constraints on the magnetic and electric dipole moments, $\mu_{\chi}$ and $d_{\chi}$, are derived separately, and the corresponding results are shown in Fig.~\ref{fig:constraints_dipole}.

We present the direct-detection constraints in both panels. These include limits from the Migdal effect together with conventional nuclear recoils from DarkSide-50 (bright red shaded region), DM-electron scattering constraints from XENON10 (pink shaded), and XENONnT (purple shaded), as well as bounds from silicon semiconductor targets in SENSEI@SNOLAB  (cyan shaded) and DAMIC-M (brown shaded). 
In addition, we show projected sensitivities from germanium semiconductor detectors, which are indicated by dashed curves without shaded regions.

The left panels in Fig.~\ref{fig:constraints_dipole} show the case with $m_{A'} = 3m_{\chi}$. 
In this case, the DM relic abundance is determined solely by the annihilation process into the SM particles since the annihilation channel $\chi \overline{\chi} \to A'A'$ is kinematically forbidden in the non-relativistic limit.
The steel-blue region is excluded in the left panels since weak dipole interactions lead to the overabundance of the DM particles.
In addition, it is expected that the amount of the anti-DM particle in the current Universe in the thermal DM case. 
Therefore, we have the cosmological and astrophysical observations constraining the late-time annihilation of a pair of DM particles into the SM particles. 
In the left panels, the cosmological constraints from CMB and BBN are shown as the orange-shaded region and the green-shaded region, respectively, while the cosmic-ray constraints are shown as the gray-shaded region.

A viable parameter space is almost closed for the case where the magnetic dipole moment $\mu_{\chi}$ is dominated, shown in the left-top panel of Fig.~\ref{fig:constraints_dipole}.
Due to no velocity suppression for the $s$-wave annihilation process, the cosmological and astrophysical observations put the strong bound on the magnetic dipole moments $\mu_{\chi}$. 
The light DM with the mass of $m_{\chi} \lesssim 6 \times 10^{-3} \, \mathrm{GeV}$ remains consistent with the direct-detection, cosmological, and astrophysical constraints. 
We note that the cosmological and astrophysical constraints can be easily relaxed if there are additional annihilation channels for DM particles.
Meanwhile, a viable parameter space remains open for a wide range of the DM mass for the case where the electric dipole moment $d_{\chi}$ is dominated, shown in the left-bottom panel of Fig.~\ref{fig:constraints_dipole}.
Since the annihilation process by the electric dipole moment $d_{\chi}$ is $p$-wave, and hence the constraints on the late-time annihilation are significantly weaker than the magnetic-dipole-dominated case.
The scattering cross section via the magnetic dipole moment Eq.~\eqref{eq:MDM_cross_section} is proportional to $q^4$, while that via the electric dipole moment Eq.~\eqref{eq:EDM_cross_section} is proportional to $m_f^2 q^2$. 
Therefore, the constraints on the dipole moments from the direct-detection experiments are more stringent for the electric dipole moment, due to the suppression of the cross section for the magnetic dipole moment by the momentum transfer. 
Since the cross section for both annihilation and scattering processes is proportional to $\mu_\chi^2 \epsilon^2$ ($d_\chi^2 \epsilon^2$), the constraints on $\mu_{\chi}$ ($d_{\chi}$) simply scales as $\mu_{\chi} \propto \epsilon^{-1}$ ($d_{\chi} \propto \epsilon^{-1}$).

Now, we turn to the cases with $m_{A'} = 0.1 m_{\chi}$, shown as the right panels in Fig.~\ref{fig:constraints_dipole}.
In this case, the dark photon is the lightest particle in the dark sector. 
In addition to the standard thermal freeze-out picture, the observed dark matter abundance can also be explained by the particle-antiparticle asymmetry, as discussed in Section~\ref{sec:DDM}. 
Motivated by this possibility, we display the cosmological constraints from relic abundance, late-time annihilation, and cosmic-ray observations using relatively transparent shading and dashed boundaries, indicating that these bounds may be relaxed or absent in asymmetric dark matter scenarios. For given DM mass, the scattering cross section is larger for lighter dark photon mass.
Thus, the direct-detection constraints in the case with $m_{A'} = 0.1 m_{\chi}$ get severer than those in the case with $m_{A'} = 3 m_{\chi}$.

For the electric dipole moment $d_{\chi}$, the strongest bound is set by the Migdal-effect limit from DarkSide-50 data: $d_\chi \lesssim 3.5\times10^{-18}\,e\,\mathrm{cm}$ at $m_\chi\simeq 5\,\text{GeV}.$ 
While for the magnetic dipole moment $\mu_{\chi}$, both DarkSide-50 and cosmological data lead to strong constraints, because the absence of velocity suppression in the s-wave channel amplifies the cosmological limits relative to the EDM case. 

Across both mass hierarchies, the magnitude of the dipole interactions is strongly constrained for DM masses above the GeV scale. 
The direct-detection experiments with the low recoil-energy threshold, such as not only the semiconductor detectors (detecting the electron recoils) but also cryogenic detectors (detecting phonon excitation, \textit{e.g.} Super-CDMS--SNOLAB~\cite{SuperCDMS:2016wui}), are essential to probe the dark dipole interactions of the dark matter particles in the MeV--GeV mass range.

We compare our dark dipole DM scenario with the electromagnetic-dipole DM scenarios (see Refs.~\cite{Pospelov:2000bq,Masso:2009mu,Banks:2010eh,Barger:2010gv,DelNobile:2014eta,Hisano:2020qkq,Ibarra:2024mpq} for the electromagnetic-dipole DM). 
The scattering cross section is suppressed by not only the kinetic mixing parameter $\epsilon$ but also the propagator of the dark photon. 
Since the relative momentum transfer is typically of $\mathcal{O}(10 \, \mathrm{keV})$ for the nuclear recoil and $\mathcal{O}(1 \, \mathrm{eV})$ for the electron recoil, the dark-photon propagator suppresses the scattering cross section by a factor of
\begin{align}
    \left(\frac{q^2}{q^2 + m_{A'}^2}\right)^2
    \simeq
    \left(\frac{q}{m_{A'}}\right)^4 \,.
\end{align}
It indicates that the scattering rate is approximately suppressed by a factor of $m_T^2 q_\mathrm{max}^4 / (3 m_{A'}^4 m_N^2)$ at low momentum transfer.
The suppression factor for the scattering rate is of $10^{-8} (m_\chi/m_{A'})^4$ at the argon detector, and hence the constraints on the dark dipole moments are weaker than those on the electromagnetic dipole moments by a factor of $10^{-4} (m_\chi/m_{A'})^2$ even for $\mathcal{O}(1)$ values of the kinetic mixing parameter.

Last but not least, we comment on the effective field theory (EFT) interpretation of the dark dipole interactions.
Since the dipole interactions are described by dimension-five operators, the EFT description is valid only below the cutoff scale $\Lambda$ of the EFT.
For instance, we may consider the UV completion of the dark dipole interactions by introducing heavy charged scalar and fermion fields, which are integrated out to generate the dipole interactions at low energy.
The dipole couplings are expected to be scaled as 
\begin{align}
    \mu_{\chi} \sim \frac{c_\mu}{\Lambda} \,, \qquad
    d_{\chi} \sim \frac{c_d}{\Lambda} \,,
\end{align}
where $c_\mu$ and $c_d$ are dimensionless coefficients including the gauge coupling and possible loop factors, and $\Lambda$ is interpreted as the mass scales of the heavy fields. 
Therefore, the constraints on the dipole moments can be translated into the constraints on the cutoff scale $\Lambda$ of the EFT.
For the $\mathcal{O}(1)$ coefficients, the constraints on the dipole moments can be translated as  
\begin{align}
    \mu_{\chi} \,, d_\chi \simeq 6 \times 10^{-15} \, e \, \mathrm{cm} \left( \frac{1 \, \mathrm{GeV}}{\Lambda} \right) \,.
\end{align}
The left panels in Fig.~\ref{fig:constraints_dipole} show that the EFT description of the dark dipole interactions is valid for the cutoff scale being close to (or larger than) the DM mass scale for DM mass below $\mathcal{O}(10) \, \mathrm{MeV}$.
For the heavier DM mass, the constraints on the dipole moments are stringent enough to require the cutoff scale $\Lambda$ to be smaller than the DM mass. 
For the larger $\epsilon$ (but would be confronted with the constraints from searching for the invisible decay signal of $A'$), the constraints on the dipole interactions from the direct detections and cosmological observations become more stringent, and the relic abundance constraint becomes weaker. 
This relaxes the EFT description with the heavier states since the smaller dipole moments are required to satisfy the constraints, and hence the cutoff scale $\Lambda$ can be larger than the DM mass scale.
Another UV realization for the dark dipole interactions is the composite DM scenario, where the DM particle is a composite state of charged constituents, which we will discuss in the next subsection.

\subsection{Constraints on the Dark Photon Mixing}

\begin{figure}[tp]
    \centering
    \includegraphics[width=0.6\textwidth]{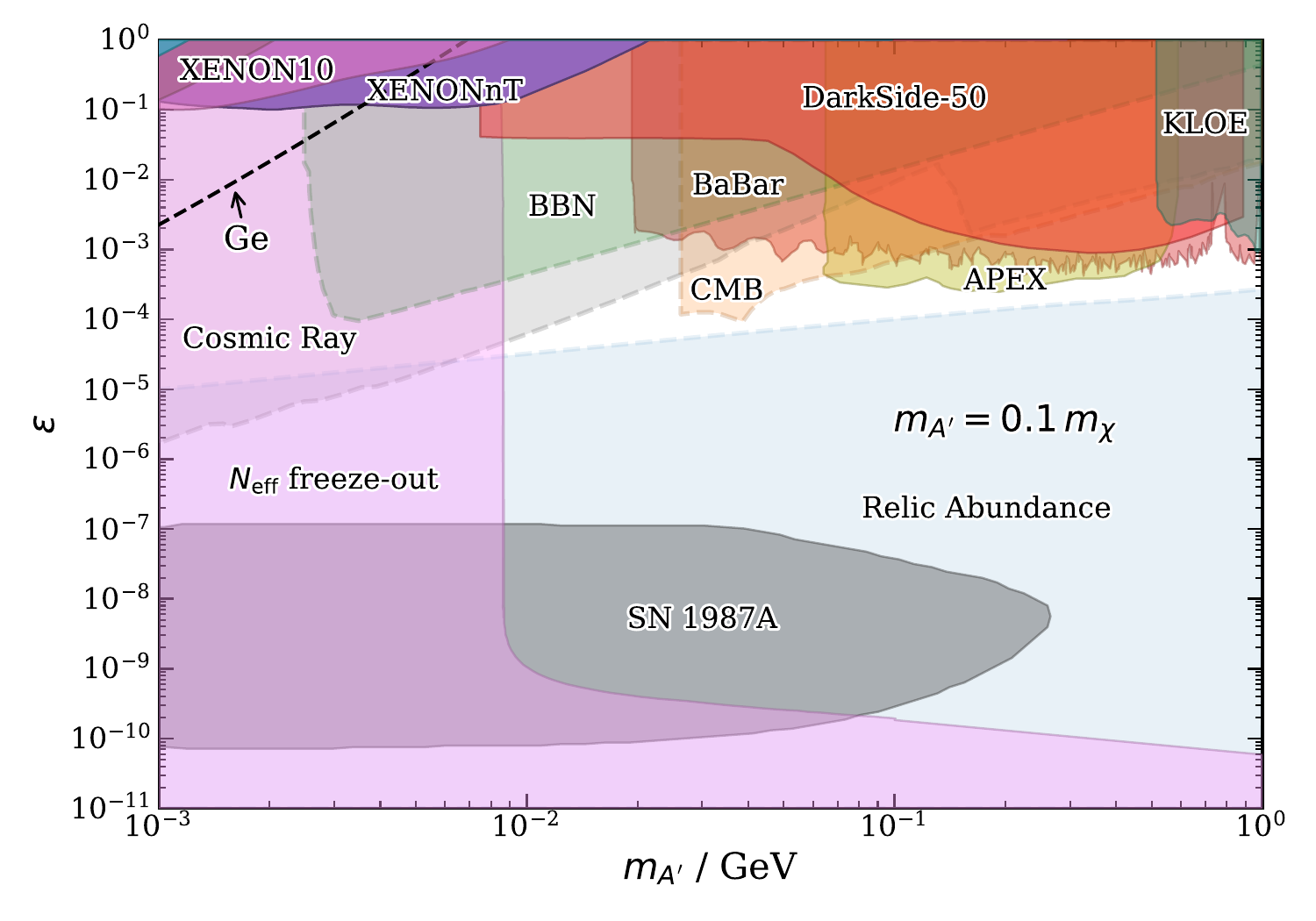}
	\caption{
    Constraints on the dark photon kinetic-mixing parameter $\epsilon$ as a function of the dark photon mass $m_{A'}$ for the case $m_{A'} = 0.1 m_{\chi}$ and $\mu_\chi = e/2m_\chi$.
    The color codes for the constraints are the same as in Fig.~\ref{fig:constraints_dipole}.
    Direct-detection limits arise from the Migdal effect (DarkSide-50: bright red), DM-electron scattering (XENON10: pink shaded, XENONnT: purple), and skipper-CCD (SENSEI: cyan), while projected sensitivities from germanium semiconductor detectors are shown as a black-dashed line. 
    Cosmological constraints include limits from $N_\mathrm{eff}$ (freeze-out scenario, light pink), relic abundance (steel-blue), BBN (green), CMB (orange) and cosmic-ray (gray).
    Again, the constraints on the annihilation processes are shown with relatively transparent colors and dashed boundaries, indicating that these bounds may be relaxed or absent in asymmetric DM scenarios.
    Collider constraints from BaBar (dark-red) and KLOE (teal), fixed-target constraints from APEX (olive) and SN1987 A constraints (dark-gray) are also included.
    Meanwhile, in terms of the visibility of the direct-detection constraints, we do not show the constraints from the long-lived particle searches from proton beam-dump and FASER, which are overlapped with the constraints from XENONnT and XENON10.
    }
	\label{fig:epsilon_constraint}
\end{figure}

Now, we consider the case where the dark dipole moments are given in analogous to the Bohr magneton of the fundamental particles. 
In this case, the scale of dipole moments are related with the DM mass, but the $U(1)_D$ coupling $e'$ is still a free parameter. 
Once we assume that the $U(1)_D$ coupling $e'$ is identical to the electromagnetic coupling, $e' = e$, we can constrain the kinetic mixing parameter $\epsilon$ between dark photon and photon for given DM mass. 
We may also think that a composite-particle realization of the DM particle, such as a neutron-like composite particle. 
\begin{align}
	\label{eq:dipole_expression}
	\mu_{\chi} = \frac{e'}{2m_{\chi}} \,, \qquad
	d_{\chi} = \frac{\theta' e'}{2m_{\chi}} \,,
\end{align}
where $\theta'$ denotes the $CP$-violating phase in the dark sector. 
$\theta'$ appeared in the electric dipole $d_{\chi}$ relies on the $CP$ violation in the dark sector and it is a free parameter, but we simply ignore the contribution $d_{\chi}$ in the following analysis.
As with the previous subsection, we may consider the two benchmark mass parameters, $m_{A'}=3m_{\chi}$ and $m_{A'}=0.1m_{\chi}$. 
However, the dipole moments scale as 
\begin{align}
    \mu_{\chi} \simeq 10^{-16} \left( \frac{1 \, \mathrm{GeV}}{m_\chi} \right) \left( \frac{e'}{e} \right)\, e \, \mathrm{cm} \,.
\end{align}
This scaling indicates that the direct-detection constraints for the dark-dipole DM is irrelevant even if we take $\epsilon \simeq 1$ for heavier dark photon case, $m_{A'} = 3m_{\chi}$, as shown in Fig.~\ref{fig:constraints_dipole}.
Hence, we omit a constraint plot for $m_{A'} = 3m_{\chi}$.
The constraints on the kinetic mixing $\epsilon$ with assuming the form of Eq.~\eqref{eq:dipole_expression} are presented in Fig.~\ref{fig:epsilon_constraint}.

The DarkSide-50 has put a constraint on the kinetic mixing $\epsilon \gtrsim 10^{-3}$ for DM mass of $1 \text{--} 10 \, \mathrm{GeV}$ ($0.1 \text{--} 1 \, \mathrm{GeV}$ dark photon), and electron-recoil experiments (XENON10 and XENONnT) have put constraints $\epsilon \gtrsim 10^{-2}$ for sub-GeV DM. 
In addition, as discussed in Section~\ref{sec:darkphoton}, there is a $N_\mathrm{eff}$ constraint on the late-time decay of dark photon, depicted as the light-pink shaded region.
Here, we assume that the dark sector is thermalized well through the dark photon portal. 
This assumption would be reasonable in this case: even though DM is neutral under $U(1)_D$, the heavier charged states or charged constituents (such as quarks for neutron) are required to generate the magnetic dipole moment for DM.
It indicates that the dark sector and the SM sector share their temperatures in the early Universe. 

We also show the collider constraints from BaBar~\cite{BaBar:2014zli} (dark-red) and KLOE~\cite{KLOE-2:2011hhj,KLOE-2:2012lii,Anastasi:2015qla,KLOE-2:2016ydq} (teal), as well as the fixed-target constraints from APEX~\cite{APEX:2011dww} (olive) and SN1987A constraints~\cite{Chang:2016ntp} (dark-gray), (see also Refs.~\cite{Caputo:2025aac,Caputo:2025avc} for updated constraints on dark photon from SN1987A).
Meanwhile, for the visibility of the direct-detection constraints of our main interests, we do not show the existing bounds from the fixed-target experiments with the proton beam-dump, such as CHARM~\cite{CHARM:1985nku,Gninenko:2012eq}, NA48/2~\cite{NA482:2015wmo}, LSND~\cite{LSND:1997vqj,Batell:2009di}, U70/$\nu$Cal~\cite{Blumlein:2011mv,Blumlein:2013cua}, and the FASER experiment at the Large Hadron Collider~\cite{FASER:2023tle}.
These constraints are overlapped with the constraints from XENONnT and XENON10 in Fig.~\ref{fig:epsilon_constraint}.
As with the right panels in Fig.~\ref{fig:constraints_dipole}, the relic-abundance, CMB, BBN, and cosmic-ray constraints are also shown in Fig.~\ref{fig:epsilon_constraint} with relatively transparent colors and dashed boundaries, indicating that these bounds may be relaxed or absent in asymmetric DM scenarios.
Consequently, in the asymmetric DM case, the direct-detection limits provide weaker constraints on the kinetic mixing parameter $\epsilon$ for $m_{A'} \gtrsim 1 \, \mathrm{MeV}$, than the collider and fixed-target constraints. 
The direct-detection constraints become more relevant for the large $U(1)_D$ coupling, $e' \gg e$, or for the case where the dipole moment somehow deviates from the scaling of the nuclear magneton.

\section{Summary and Conclusion \label{sec:Conc}}


In this work, we have investigated a fermionic DM that interacts with the SM particles exclusively through electric and magnetic dipole couplings to a massive dark photon. 
By assuming that the DM particle $\chi$ is neutral under $U(1)_D$, the leading interactions arise only through dipole operators, providing an alternative to dark matter models with current-mediated interactions involving a massive dark photon.
Such a structure naturally appears in dark sectors with confining gauge dynamics, where composite neutral states can acquire dipole moments analogous to the neutron.

The dipole nature of the DM-SM interactions leads to the different momentum transfer dependence of the scattering and annihilation processes. 
We consider three kinds of phenomenological constraints to explore the viable parameter space of this framework: relic abundance, cosmological/astrophysical constraints, and direct-detection constraints.  
Regarding the DM relic abundance, the dominant freeze-out processes may involve annihilation either into the SM particles or into pairs of dark photons, depending on the mass hierarchy between the DM and the dark photon. 
When the dark photon is heavier than the DM, only the SM final states is allowed for the DM annihilation process.
Hence, the relic abundance is sensitive to both the dipole interaction and kinetic mixing between the dark photon and the SM photon. 
Meanwhile, when the dark photon is lighter, the annihilation into dark photons determines the DM relic abundance. 
The relic abundance may place the lower bound only on the dipole interactions. 
For this mass spectrum, we can also explain the relic abundance as the asymmetric DM scenario, rather than the thermal freeze-out scenario. 
In this case, there is no constraint from the relic abundance (and no or weak constraints from cosmological and astrophysical observations).

Cosmological/astrophysical constraints would drastically change depending whether the magnetic dipole or the electric dipole dominates the annihilation processes. 
The magnetic dipole interaction leads to the $s$-wave annihilation, while the electric dipole interaction leads to the $p$-wave annihilation. 
Hence, depending on the kinds of dipole interactions, the CMB, BBN, and cosmic-ray constraints from the late-time energy injection into the electromagnetic plasma would change: they are quite strong for magnetic dipole interactions ($s$-wave), while they are weak for electric dipole interactions ($p$-wave).
Together withe the relic abundance requirement, these cosmological and astrophysical observations significantly restrict the magnetic dipole interactions when the dark photon is heavier than the DM.
When the dark photon is lighter than the DM, these constraints can be avoided in the asymmetric DM scenario. 
On the other hand, the dark photon being the lightest dark-sector particle leads to entropy transfer after neutrino decoupling, which reduces the effective number of neutrino species $N_\mathrm{eff}$.

The direct-detection constraints, which are the main focus of this work, arise from various experiments: in this study, we have incorporated the relevant methods, nuclear recoils (including the Migdal effect), DM-electron scattering, and semiconductor targets.
Each method requires the different sizes of momentum transfer and typical energy scales of reaction processes, and hence each method provides the constraints in the different DM mass ranges.
The nuclear-recoil constraints are sensitive to the DM mass in the GeV range and above, and the constraints extend to the sub-GeV mass range by including the Migdal effect.
The direct detection experiments utilizing the electron recoils (in particular, using the LXe detector, such as XENON10 and XENONnT) are sensitive to the DM mass in 10\,MeV to GeV range. 
The direct detection experiments utilizing semiconductor targets (SENSEI and DAMIC-M) extend sensitivity even at the DM mass below 10\,MeV thanks to the eV-scale band gap in the semiconductor targets.

Figures~\ref{fig:constraints_dipole} and  \ref{fig:epsilon_constraint} summarize the constraints on the dark dipole moments and the dark photon kinetic mixing, respectively.
For the case where the dark photon is heavier than the DM, namely the thermal freeze-out scenario, the combination of relic abundance and late-time cosmology severely constrains the magnetic dipole interactions, leaving only a small region with $m_{\chi} \lesssim 6 \times 10^{-3}$ GeV viable.
Meanwhile, thanks to the velocity suppression of $p$-wave annihilation, the electric dipole interactions remain less constrained by the cosmological and astrophysical observations, with allowed magnitudes around $d_{\chi} \sim 10^{-12} \, e \, \mathrm{cm}$ for sub-GeV DM.
In contrast, when the dark photon is lightest in the dark sector, the relic abundance constraints and the late-time cosmology constraints can be avoided in the asymmetric DM scenario, leading to a significant expansion of the allowed parameter space.
In this case, the direct-detection experiments get more relevant to constrain the dark dipole interactions, with the DarkSide-50 Migdal bound providing the strongest limits near a few GeV for both EDM and MDM scenarios.
In all cases, semiconductor projections indicate that next-generation solid-state experiments could further improve sensitivity in the MeV--sub-GeV mass range.

Last but not least, we show the constraints in terms of the dark photon kinetic mixing parameter $\epsilon$ by assuming that the dark dipole moments scale like the nuclear magneton, presented in Fig.~\ref{fig:epsilon_constraint}.
In this case, the direct-detection constraints are generally weaker than the existing collider and fixed-target constraints for $m_{A'} \gtrsim 1 \, \mathrm{MeV}$ and $N_\mathrm{eff}$, which comes from the late-time entropy transfer from the dark photon decay.
Notably, substantial regions of parameter space remain viable in the sub-GeV $m_{A'}$ region, particularly where entropy-transfer constraints weaken.

There remain several avenues for extending the present study. 
Future low-threshold semiconductor detectors (not only electron recoils but also detecting quasi-particle excitations) are expected to substantially enhance sensitivity to light dark matter.
Not only the DM direct detections, but also the collider experiments, the fixed-target experiments, and cosmological/astrophysical observations will improve the sensitivity to the dark photon and dark dipole interactions in the near future. 


\section*{Acknowledgements}
T.K thanks Ayuki Kamada for fruitful discussions.
The work of T.K. is supported in part by the National Science Foundation of China (NSFC) under Grant Nos.~11675002, 11635001, 11725520, 12235001, and the NSFC Research Fund for International Scientists Grant No.~12250410248.

\appendix
\section{Ionization Form Factor \label{app:ionization}}

In Section~\ref{sec:Migdal_effect}, we have discussed the Migdal effect as a probe of dark dipole interactions. 
The corresponding event rate is determined by the transition probabilities $p^d_{q_e}$ and $p^c_{q_e}$, given in Eq.~\eqref{eq:Migdal_transition}: $p^d_{q_e}$ describes bound-state excitations, and $p^c_{q_e}$ describes ionization into scattering states. 
In the following, we derive the explicit expressions for transition probabilities.  
These transition probabilities are explicitly derived in Ref.~\cite{Ibe:2017yqa}.
\begin{align}
    \label{eq:pd_expression}
    p_{q_e}^{d}(n \ell \to n' \ell')&
    = \frac{2(2\ell'+1)-\omega_{n',\ell'}}{2(2\ell'+1)}\frac{\omega_{n,\ell}}{2(2\ell+1)} \sum_{\kappa,\kappa',m,m'} \delta_{\ell,|\kappa+1/2|-1/2} \delta_{\ell',|\kappa'+1/2|-1/2} \nonumber \\ 
    &\times | z_{\vb*{q}_e}(E_{n',\kappa'},\kappa',m'|E_{n,\kappa},\kappa,m)|^2 \,, \\
    \label{eq:pc_expression}
    \frac{\mathrm{d}}{\mathrm{d}E_e}p_{q_e}^{c}(n\ell\to E_e)&=\frac{\omega_{n,\ell}}{2(2\ell+1)} \sum_{\kappa,\kappa',m,m'} \delta_{\ell,|\kappa+1/2|-1/2} | z_{\vb*{q}_e}(E_e,\kappa',m'|E_{n,\kappa},\kappa,m)|^2,
\end{align}
where $\kappa^{(\prime)}$ and $m^{(\prime)}$ are the labels of the spherical spinor harmonics: $m$ denotes the magnetic quantum number, and the relation of $\kappa$ to the total angular momentum $j$ and orbital angular momentum $\ell$ is given by $\kappa = \mp (j+1/2)$ for $j = \ell \pm 1/2$. 
Namely, the absolute value of $\kappa$ is related to the total angular momentum $j$, and the sign of $\kappa$ determines the relative sign of $j - \ell$. 
$\omega_{n,\ell}$ is the electron occupation number of quantum states $(n,\ell)$, while $E_{n,\kappa}$ represents the binding energy for bound electrons $(n,\kappa)$, given by
\begin{align}
    \label{eq:binding_energy_for_Dirca_electron}
    E_{n,\kappa}=\frac{m_e}{\sqrt{1+\frac{\alpha^2Z^2}{(\gamma+n-|\kappa|)^2}}} \,, \quad 
    \gamma=\sqrt{\kappa^2-\alpha^2Z^2} \,,
\end{align}
with $Z$ being the atomic number. 
The transition amplitude for an atomic state labeled by $(E_{n,\kappa},\kappa,m)$ to a state $(E_{n',\kappa'},\kappa',m')$, denoted by $z_{\vb*{q}_e}(E_{n',\kappa'},\kappa',m'|E_{n,\kappa},\kappa,m)$ in Eqs.~\eqref{eq:pd_expression} and \eqref{eq:pc_expression}, is obtained as:
\begin{align}
    \label{eq:transition_amp}
    z_{\vb*{q}_e}(E',\kappa',m'|E,\kappa,m) 
    = - i q_e \int_{0}^{\infty}\mathrm{d}rr\left[ P_{E',\kappa'}(r)P_{E,\kappa}(r)+Q_{E',\kappa'}(r)Q_{E,\kappa}(r)\right] \nonumber \\
    \times \int_{0}^{2\pi} \mathrm{d}\varphi \int_{0}^{\pi}\mathrm{d}\theta\sin \theta \cos \theta \Omega^{\dagger}_{\kappa',m'}(\theta,\varphi)\Omega_{\kappa,m}(\theta,\varphi) \,,
\end{align}
where $\Omega_{\kappa,m}(\theta,\varphi)$ are the spherical spinor harmonics, given by
\begin{align}
    \label{eq:spherical_spinor}
    \Omega_{\kappa,m}(\theta,\varphi) & = 
    \begin{pmatrix}
        -\sqrt{\frac{\kappa-m+1/2}{2\kappa+1}}Y_{\kappa,m-1/2}(\theta,\varphi)\\
        \sqrt{\frac{\kappa+m+1/2}{2\kappa+1}}Y_{\kappa,m+1/2}(\theta,\varphi)\\
    \end{pmatrix} \,, 
    & & \kappa \geq 0 \,, \\ 
    \Omega_{\kappa,m}(\theta,\varphi) & =
    \begin{pmatrix}
        \sqrt{\frac{\kappa-m+1/2}{2\kappa+1}}Y_{-\kappa-1,m-1/2}(\theta,\varphi)\\
        \sqrt{\frac{\kappa+m+1/2}{2\kappa+1}}Y_{-\kappa-1,m+1/2}(\theta,\varphi)\\
    \end{pmatrix} \,, 
    & & \kappa < 0 \,.
\end{align}
Here, $Y_{\ell m}(\theta, \varphi)$ denotes the spherical harmonics. 
Once we assume the hydrogen-like atoms (namely, we ignore the electron-electron interactions and screening effects), the radial functions for the Dirac equations can be analytically expressed.
In this case, the radial functions for the Dirac equations $P_{E,\kappa}(r)$ and $Q_{E,\kappa}(r)$ in Eq.~\eqref{eq:transition_amp} are given by
\begin{align}
    \label{eq:radial_Dirca_function_P}
    P_{E,\kappa}(r)&=N_{E,\kappa}\sqrt{1+\frac{E}{m_e}}e^{-\lambda r}(2\lambda r)^{\gamma}\left[\left(-\kappa+\frac{Zm_e}{\lambda}\right)F(a,b,2\lambda r)+aF(a+1,b,2\lambda r)\right], \\
    \label{eq:radial_Dirca_function_Q}
    Q_{E,\kappa}(r)&=N_{E,\kappa}\sqrt{1-\frac{E}{m_e}}e^{-\lambda r}(2\lambda r)^{\gamma}\left[\left(-\kappa+\frac{Zm_e}{\lambda}\right)F(a,b,2\lambda r)-aF(a+1,b,2\lambda r)\right],
\end{align}
where, $\lambda = \sqrt{m_e^2-E^2}$, $a =\gamma-ZE/\lambda$, $b = 2\gamma+1$, $N_{E,\kappa}$ is the normalization factor, and $F(a,b,x)$ denotes the confluent hypergeometric function. 
With all these components, $p_{q_e}^{d}$ and $\mathrm{d}p_{q_e}^{c}/\mathrm{d}E_e$ can be explicitly computed.

\section{Dark Matter-Nucleus Cross Section \label{app:DM-Nucleus}}

In this appendix, we provide a detailed derivation of the DM-nucleus differential scattering cross section induced by dipole interactions, corresponding to Eq.~\eqref{eq:diff_cross_section_nucleus} in the text.

\subsection{General Procedure}
\label{sec:general_procedure}
In this subsection, we outline the general framework for deriving the DM-nucleus differential cross section $\mathrm{d}\sigma_T/\mathrm{d}E_R$, closely following the formalism developed in Ref.~\cite{DelNobile:2021wmp}. 
In the non-relativistic (NR) limit, the differential cross section can be expressed as
\begin{equation}
\label{eq:expression_NR}
    \frac{\mathrm{d}\sigma}{\mathrm{d}E_{R}}\overset{\mathrm{NR}}{=}
    \frac{1}{32\pi}\frac{1}{m_{\chi}^2 m_T}\frac{1}{v^2} \overline{|\mathcal{M}|^2},
\end{equation}
where $m_\chi$ and $m_T$ denote the masses of the DM particle and the target nucleus, respectively, and $v$ is the relative velocity of the DM particle to the target nucleus.
$\overline{|\mathcal{M}|^2}$ denotes the spin-averaged squared amplitude for the scattering process. 
The central task is therefore to compute the squared amplitude in the NR limit.

To this end, one should first evaluate the DM-nucleon scattering amplitude $\mathcal{M}_N$ in a relativistic manner and subsequently perform a NR expansion of the scattering amplitude. 
The Dirac spinor in the Weyl basis is expanded in the NR limit as follows.
\begin{align}
    \label{eq:spinor_non}
    u^{s}(p)=
    \begin{pmatrix} 
        \sqrt{p\cdot \sigma} \xi^{s}\\ 
        \sqrt{p\cdot \overline{\sigma}}\xi^{s} 
    \end{pmatrix}
    \overset{\mathrm{NR}}{=}
    \frac{1}{\sqrt{4m}}
    \begin{pmatrix} 
        (2mI_2 -\bm{p\cdot\sigma}) \xi^{s}\\ 
        (2mI_2+\bm{p\cdot \sigma}) \xi^{s}  
    \end{pmatrix} \,,
\end{align}
where $m$ denotes the mass of the Dirac fermion, and $\xi^{s}$ denotes the two-component spinor: $\xi^{s}=(1,0)^T$ corresponds to spin up along the $z$ axis, while $\xi^{s}=(0,1)^T$ corresponds to spin down.
The Pauli matrices are represented by $\bm{\sigma}$, and $I_2$ denotes the $2\times 2$ identity matrix: 
\begin{align}
    I_2 =
    \begin{pmatrix}
        1 & 0 \\
        0 & 1 
    \end{pmatrix}\,, \quad
    \sigma^1 = & 
    \begin{pmatrix}
        0 & 1 \\
        1 & 0 
    \end{pmatrix} \,, \quad 
    \sigma^2 = 
    \begin{pmatrix}
        0 & -i \\
        i & 0 
    \end{pmatrix} \,, \quad 
    \sigma^{3} = 
    \begin{pmatrix}
        1 & 0 \\
        0 & -1 
    \end{pmatrix} \,.
\end{align}
The gamma matrices and other relevant quantities in the Weyl basis are defined by
\begin{align}
    \label{eq:gamma_matrix}
    \gamma^{\mu} =
    \begin{pmatrix}
        0 & \sigma^{\mu} \\
        \overline{\sigma}^{\mu} & 0 
    \end{pmatrix} \,, \quad 
    \gamma^{5} & =
    \begin{pmatrix}
        -I_2 & 0 \\
        0 & I_2 
    \end{pmatrix}\,, \quad 
    \sigma^{\mu\nu} =
    \begin{pmatrix}
        \sigma_2^{\mu\nu} & 0 \\
        0 & \overline{\sigma}_2^{\mu\nu} 
    \end{pmatrix} \,, \\ 
    \sigma^{\mu} = (I_2, \bm{\sigma})^{T} \,, \quad
    \overline{\sigma}^{\mu} = (I_2,-\bm{\sigma})^{T} \,, \quad
    & \sigma_2^{\mu\nu} = \frac{i}{2}(\sigma^{\mu}\overline{\sigma}^{\nu}-\sigma^{\nu}\overline{\sigma}^{\mu}) \,, \quad
    \overline{\sigma}^{\mu\nu}_{2}=\frac{i}{2}(\overline{\sigma}^{\mu}\sigma^{\nu}-\overline{\sigma}^{\nu}\sigma^{\mu})\, .
\end{align}
For later convenience, we also introduce the kinematic variables.
Let us define the four-momenta of the initial (final) DM particle and nucleon by $p^\mu$ ($p'^\mu$) and $k^\mu$ ($k'^\mu$), respectively.
In the NR limit, the sum of the initial and final four-momenta, and the momentum transfer are given by
\begin{align}
    \label{eq:kin_quantity}
    P^{\mu} & \equiv p^{\mu} + p'^{\mu}
    \overset{\mathrm{NR}}{=}
    \begin{pmatrix}
        2m_{\chi} \\
        \bm{p}+\bm{p'}
    \end{pmatrix} 
    \equiv
    \begin{pmatrix}
        2m_{\chi} \\
        \bm{P}
    \end{pmatrix} \,, \\
    K^{\mu} & \equiv k^{\mu} + k'^{\mu}
    \overset{\mathrm{NR}}{=}
    \begin{pmatrix}
        2m_{N} \\
        \bm{k}+\bm{k'}
    \end{pmatrix} \equiv
    \begin{pmatrix}
        2m_N \\
        \bm{K}
    \end{pmatrix} \,, \\
    q^{\mu} & \equiv p^{\mu}-p'^{\mu} 
    \overset{\mathrm{NR}}{=}
    \begin{pmatrix}
        \frac{\bm{K\cdot (p-p')}}{2m_{N}} \\
        \bm{p-p'}
    \end{pmatrix} 
    \equiv 
    \begin{pmatrix}
        \frac{\bm{K\cdot q}}{2m_{N}} \\ \notag
        \bm{q}
    \end{pmatrix} \,.
\end{align}
We define the identity and spin operators for the DM particle and the nucleon as follows.
\begin{align}
    \mathcal{I}_{\chi} & \equiv \xi^{s'\dagger}_{\chi}\xi^{s}_{\chi} \,, \quad 
    \bm{s}_{\chi} \equiv \xi^{s'\dagger}_{\chi}\frac{\bm{\sigma}}{2}\xi^{s}_{\chi} \,,
    \\
    \mathcal{I}_{N} & \equiv \xi^{s'\dagger}_{N}\xi^{s}_{N}\,, \quad 
    \bm{s}_{N} \equiv \xi^{s'\dagger}_N \frac{\bm{\sigma}}{2}\xi^{s}_N\,.
\end{align}
Here, $\xi^{s}_{\chi}$ and $\xi^{s}_N$ are the two-component spinors for the DM particle and the nucleon, respectively.
The relative velocity of the DM particle to the nucleon is defined as
\begin{align}
    \bm{v} \equiv \bm{v}_\mathrm{DM} - \bm{v}_N
    = \frac{\bm{p}}{m_{\chi}}&-\frac{\bm{k}}{m_{N}} \,, \quad 
    \bm{v}_N^{\perp} \equiv \frac{1}{2}(\bm{v}_\mathrm{DM}+\bm{v'}_\mathrm{DM}-\bm{v}_{N}-\bm{v'}_N) \,.
\end{align}
With these definitions, the DM-nucleon amplitude in the NR limit can always be decomposed as~\cite{DelNobile:2021wmp}
\begin{equation}
    \label{eq:M_N_non}
    \mathcal{M}_{N}\overset{\mathrm{NR}}{=}
    \sum_{i} f^{N}_{i}(q^2,v_{N}^{\perp^2}) \mathcal{O}_{i}^{N} \,,
\end{equation}
where the coefficients $f_i^N$ are functions of $q^2$ and $v_N^{\perp^2}$, and $\mathcal{O}_i^N$ are the NR operators constructed from $\bm{q}$, $\bm{v}_N^{\perp}$, $\bm{s}_{\chi}$, and $\bm{s}_N$. 
For spin-$1/2$ DM, a complete basis consists of 16 such operators~\cite{DelNobile:2021wmp}. 
In this work, only the following subset is relevant:
\begin{align}
    \mathcal{O}^{N}_1 & = 1 \,, \quad 
    \mathcal{O}^{N}_4 = \bm{s}_{\chi}\bm{\cdot}\bm{s}_N \,, \nonumber  \\
    \mathcal{O}^{N}_5 & = i\bm{s}_{\chi}\bm{\cdot}(\bm{q\times}\bm{v}_{N}^{\perp})\,, \quad 
    \mathcal{O}^{N}_6 = (\bm{s}_{\chi}\bm{\cdot}\bm{q})(\bm{s}_{N}\cdot\bm{q}) \,, \quad 
    \mathcal{O}^{N}_{11} = i\bm{s}_{\chi}\bm{\cdot}\bm{q} \,.
    \label{eq:NR_operator}
\end{align}
Once the NR amplitude $\mathcal{M}_N$ for the DM-nucleon scattering is determined, the unpolarized squared amplitude can be written in the compact form~\cite{DelNobile:2021wmp}
\begin{equation}
    \label{eq:M_NR_final}
    \overline{|\mathcal{M}|^2}
    \overset{\mathrm{NR}}{=} 
    \frac{m_{T}^{2}}{m_{N}^2} \sum_{i,j}\sum_{N,N'=p,n} f_{i}^{N}(q^2,v_{N}^{\perp^2}) f_{j}^{N'}(q^2,v_{N}^{\perp^2}) F_{i,j}^{(N,N')}(q^2,v_{N}^{\perp^2}) \,,
\end{equation}
where $F_{i,j}^{(N,N')}$ encodes the correlations between $\mathcal{O}_{i}^{N}$ and $\mathcal{O}_j^{N'}$. 
For the NR operators listed above, the non-vanishing $F_{i,j}^{(N,N')}$ are~\cite{DelNobile:2021wmp}
\begin{align}
    \label{eq:NR_form_factor}
    F_{1,1}^{(N,N')} & = F_{M}^{(N,N')} \,, 
    F_{4,4}^{(N,N')} = \frac{1}{16} \left(F_{\Sigma'}^{(N,N')}+F_{\Sigma''}^{(N,N')}\right) \,, \\ 
    F_{6,6}^{(N,N')} & = \frac{q^4}{16} F_{\Sigma''}^{(N,N')}\,, \quad 
    F_{11,11}^{(N,N')} = \frac{q^2}{4}F_{M}^{(N,N')} \,, \quad 
    F_{4,5}^{(N,N')} = -\frac{q^2}{8m_N}F_{\Sigma',\Delta}^{(N,N')} \,, \\
    F_{4,6}^{(N,N')} & = \frac{q^2}{16}F_{\Sigma''}^{(N,N')} \,, \quad 
    F_{5,5}^{(N,N')} = \frac{q^2}{4} \left(v_{N}^{\perp^2}F_{M}^{(N,N')} + \frac{q^2}{m_N^2}F_{\Delta}^{(N,N')}\right) \,.
\end{align}
Here, $F_{M}^{(N,N')}$, $F_{\Sigma'}^{(N,N')}$, $F_{\Sigma''}^{(N,N')}$, $F_{\Delta}^{(N,N')}$, and $F_{\Sigma',\Delta}^{(N,N')}$ are the nuclear response functions, which are numerically computed in Ref.~\cite{Anand:2013yka}.
Substituting all ingredients into Eq.~\eqref{eq:expression_NR} then yields the desired differential cross section.

\subsection{DM-Nucleus Cross Section with Dipole Interactions \label{sec:DM_sigma_cross_section}}

We now apply the general procedure outlined above to the specific case of electric and magnetic dipole interactions mediated by a massive dark photon.

The starting point is the DM-nucleon scattering amplitude. 
For electric and magnetic dipole interactions, the amplitudes take the form
\begin{align}
    \label{eq:nucleon_amplitude_EDM}
    \mathcal{M}_{N}^\mathrm{EDM} & 
    = - i \epsilon e \langle \chi'|J^{\mu}_\mathrm{EDM}|\chi \rangle\left(\frac{ig_{\mu\nu}}{q^2+m^2_{A'}}\right)\langle N'|J^{\nu}_\mathrm{EM}|N\rangle \,, \\
    \label{eq:nucleon_amplitude_MDM}
    \mathcal{M}_{N}^\mathrm{MDM} & 
    = - i \epsilon e \langle \chi'|J^{\mu}_\mathrm{MDM}|\chi \rangle\left(\frac{ig_{\mu\nu}}{q^2+m^2_{A'}}\right)\langle N'|J^{\nu}_\mathrm{EM}|N\rangle \, .
\end{align}
where the relevant currents are
\begin{align}
    \label{eq:current_full}
    J^{\mu}_\mathrm{EM} = 
    -Q_{N}\overline{N}\gamma^{\mu}N \,, \quad 
    J^{\mu}_{\chi, \mathrm{EDM}} = -d_{\chi}\partial_{\nu}(\overline{\chi}i\sigma^{\mu\nu}\gamma^{5}\chi) \,, \quad 
    J^{\mu}_{\chi, \mathrm{MDM}} = -\mu_{\chi}\partial_{\nu}(\overline{\chi}\sigma^{\mu\nu}\chi) \,,
\end{align}
with $Q_N=1$ $(0)$ for protons (neutrons). 
Using the NR expansions introduced in Eqs.~\eqref{eq:spinor_non} and \eqref{eq:gamma_matrix}, the matrix elements are reduced to
\begin{align}
    \label{eq:matrix_element_NR_EM}
    \langle N'|J^{\nu}_\mathrm{EM}|N \rangle 
    & = \overline{u}_{N'} \left( F_{1}^{N} \frac{K^{\mu}}{2m_{N}} + (F_1^{N}+F_2^{N}) \frac{i\sigma^{\mu \nu}q_{\nu}}{2m_{N}}\right) u_{N} 
    \overset{\mathrm{NR}}{=}
    \begin{pmatrix}
        2m_{N} \mathcal{I}_N Q_{N} \\
        \bm{K} \mathcal{I}_N Q_{N} + i g_{N} \bm{s}_{N}\bm{\times}\bm{q}
    \end{pmatrix}\, , \\
    \label{eq:matrix_element_NR_EDM}
    \langle \chi'|J^{\mu}_\mathrm{EDM}|\chi \rangle 
    & = d_{\chi} \overline{u}_{\chi'} i\sigma^{\mu\nu} \gamma^{5} u_{\chi} i q_{\nu} 
    \overset{\mathrm{NR}}{=} d_{\chi} 
    \begin{pmatrix}
        4i m_{\chi} \bm{s}_{\chi} \bm{\cdot} \bm{q} \\
        \bm{0}
    \end{pmatrix} \,, \\
    \label{eq:matrix_element_NR_MDM}
    \langle \chi'|J^{\mu}_\mathrm{MDM}|\chi \rangle 
    & = \mu_{\chi} \overline{u}_{\chi'}\sigma^{\mu\nu}u_{\chi}iq_{\nu}
    \overset{\mathrm{NR}}{=} \mu_{\chi}
    \begin{pmatrix}
        q^2\mathcal{I}_{\chi} + 2i\bm{s}_{\chi}\bm{\cdot}(\bm{q\times P}) \\
        4 i m_{\chi} (\bm{s}_{\chi}\bm{\times q})
    \end{pmatrix} \,.
\end{align}
Here, $F_1^N$ and $F_2^N$ are the nucleon electromagnetic form factors, and $g_N$ denotes the Landé $g$-factor with $g_p=5.59$ and $g_n=-3.83$. 
Substituting these expressions into the matrix elements, Eqs.~\eqref{eq:nucleon_amplitude_EDM} and \eqref{eq:nucleon_amplitude_MDM}, we obtain the matrix elements in the NR limit. 
As for the electric dipole interaction, we have
\begin{align}
    \label{eq:amplitude_NR_EDM}
    \mathcal{M}_{N}^\mathrm{EDM} 
    & \overset{\mathrm{NR}}{=} 
    \frac{8 \epsilon e}{q^2+m_{A'}^2} d_{\chi} m_{N} m_{\chi} Q_N (i\bm{s}_{\chi}\bm{\cdot}\bm{q}) 
     =
     \frac{8\epsilon e}{q^2+m_{A'}^2} d_{\chi} m_{N} m_{\chi} Q_{N} \mathcal{O}^{N}_{11} \,.
\end{align}
Here, we use the definitions of the NR operators in Eq.~\eqref{eq:NR_operator}.
As for the magnetic dipole interaction, we have
\begin{align}
    \mathcal{M}_{N}^\mathrm{MDM} 
    & \overset{\mathrm{NR}}{=} 
    \frac{2 \epsilon e \mu_{\chi}}{q^2+m_{A'}^2} 
    \left[
        m_{N} Q_{N} \mathcal{I}_{\chi} I_{N}q^2 
        + 4 i m_{\chi} m_{N} Q_{N} \mathcal{I}_{N} \bm{s}_{\chi} \bm{\cdot} (\bm{q\times v}_{N}^{\perp}) 
        + 2m_{\chi}g_{N}(\bm{s}_{\chi} \bm{\times q}) \bm{\cdot} (\bm{s}_{N}\bm{\times q})
    \right] \nonumber \\ 
    & = \frac{2\epsilon e q^2 \mu_{\chi}}{q^2+m_{A'}^2} 
    \left[ 
        m_N Q_N + 4 i \frac{m_{\chi}m_{N}}{q^2} Q_N \bm{s}_{\chi} \bm{\cdot} (\bm{q\times v}_{N}^{\perp}) + 2m_{\chi} g_{N} 
        \left(\bm{s}_{\chi}\bm{\cdot}\bm{s}_{N} - \frac{(\bm{s}_{\chi}\bm{\cdot q})(\bm{s}_{N}\bm{\cdot q})}{q^2} \right) 
    \right] \,,
    \label{eq:amplitude_NR_MDM}
\end{align}
where we use the identity $(\bm{s}_{\chi}\bm{\times q})\bm{\cdot}(\bm{s}_{N}\bm{\times q}) = q^2 \bm{s}_{\chi}\bm{\cdot}\bm{s}_{N}-(\bm{s}_{\chi}\bm{\cdot q})(\bm{s}_{N}\bm{\cdot q})$ in the second equality. 
Matching these expressions onto the operator basis in Eqs.~\eqref{eq:M_N_non} and \eqref{eq:NR_operator} gives
\begin{align}
    \mathcal{M}_{N}^\mathrm{MDM} & 
    \overset{\mathrm{NR}}{=} 
    \frac{2\epsilon e q^2 \mu_{\chi}}{q^2+m_{A'}^2} \left[m_{N}Q_{N}\mathcal{O}_{1}^{N} 
    + 4\frac{m_{\chi}m_N}{q^2} Q_N \mathcal{O}_5^{N} + 2 m_{\chi} g_{N} \left(\mathcal{O}_{4}^{N}-\frac{\mathcal{O}_{6}^{N}}{q^2}\right)
    \right] \,.
\end{align}
The corresponding squared amplitudes follow directly from Eqs.~\eqref{eq:M_NR_final} and \eqref{eq:NR_form_factor}:
\begin{align}
    \label{eq:amp_square_EDM_NR}
    \overline{|\mathcal{M}^\mathrm{EDM}|^2} \overset{\mathrm{NR}}{=} & 
    64\pi \epsilon^2 \alpha \left( \frac{q^2}{q^2+m_{A'}^2} \right)^2 m_{T}^2 m_{\chi}^2 d_{\chi}^2 \frac{1}{q^2}F_{M}^{(p,p)} \,, \\ 
    \overline{|\mathcal{M}^\mathrm{MDM}|^2} \overset{\mathrm{NR}}{=} & 
    16\pi \epsilon^2 \alpha \left( \frac{q^2}{q^2+m_{A'}^2} \right)^2 \frac{m_{\chi}^2m_T^2}{m_{N}^2} \mu_{\chi}^2 \bigg[ \left(\frac{1}{m_{\chi}^2}-\frac{1}{\mu_{T}^2} + 4\frac{v^2}{q^2} \right) m_{N}^2 F_{M}^{(p,p)} \nonumber \\
    & + 4F_{\Delta}^{(p,p)} - 2 \sum_{N=p,n} g_{N} F_{\Sigma',\Delta}^{(N,p)} + \frac{1}{4}\sum_{N,N'=p,n}g_{N}g_{N'}F_{\Sigma'}^{(N,N')}\bigg] \,.
\label{eq:amp_square_MDM_NR}
\end{align}
Here, we have used $Q_p=1$ and $Q_n=0$.
Finally, substituting these results into Eq.~\eqref{eq:expression_NR}, we obtain
\begin{align}
    \label{eq:final_result_NR}
    \frac{\mathrm{d}\sigma}{\mathrm{d}E_{R}} 
    & \overset{\mathrm{NR}}{=}
    \frac{1}{32\pi} \frac{1}{m_{\chi}^2m_T} \frac{1}{v^2} \left(\overline{|\mathcal{M}^\mathrm{EDM}|^2}+\overline{|\mathcal{M}^\mathrm{MDM}|^2}\right) \\
    & = \frac{\epsilon^2 \alpha}{2} \left( \frac{q^2}{q^2 + m^2_{A'}}\right)^2 \frac{m_{T}}{m^2_{N}} \frac{1}{v^2} \left\{ \mu^2_{\chi} \left[\left( \frac{1}{m^2_\chi} - \frac{1}{\mu^2_{T\chi}} + \frac{4 v^2}{q^2}\right)m^2_N F^{(p,p)}_M + 4 F^{(p,p)}_\Delta \right. \right. \nonumber \\ 
    & \qquad \left. \left. + \frac{1}{4} \sum_{N,N'=p,n} g_N g_{N'}F^{(N, N')}_{\Sigma'} - \sum_{N=p,n} 2 g_N F^{(N,p)}_{\Sigma', \Delta}\right] + 4 d^2_{\chi} \frac{m^2_N}{q^2} F^{(p,p)}_M \right\} \,,
\end{align}
which exactly reproduces Eq.~\eqref{eq:diff_cross_section_nucleus} in the text.

\bibliographystyle{utphys}
\bibliography{ref}
\end{document}